\documentclass[
 reprint,
 amsmath,amssymb,
 prx,
 aps,
]{revtex4-2}

\usepackage{graphicx}
\usepackage{siunitx}
\usepackage{mathtools}
\usepackage{color}
\usepackage{braket}
\usepackage{dcolumn}
\usepackage[normalem]{ulem}
\usepackage{pdfpages}

\usepackage[english]{babel}
\usepackage{blindtext}
\usepackage{lipsum}  
\usepackage{comment}
\usepackage{amsthm}
\usepackage{tabularx}
\usepackage{bm}
\usepackage{dsfont}
\usepackage{bbold}
\usepackage{ulem}
\usepackage{pgffor}
\usepackage{mwe,tikz}
\usepackage[percent]{overpic}
\usepackage{todonotes}
\usepackage{algorithmic}
\usepackage{uri}

\DeclareUnicodeCharacter{0308}{\"{a}}

\DeclareUnicodeCharacter{0301}{\"{o}}

\setcitestyle{super,open={},close={}}
\DeclareSIUnit\angstrom{\protect \text {Å}}

\usepackage{epstopdf}

\DeclareSIUnit{\atomicunit}{a.u.}
\DeclareSIUnit{\nanometer}{nm}

\makeatletter
\def\@fnsymbol#1{%
  \ensuremath{%
    \ifcase#1
    \or *
    \or \dagger
    \or \ddagger
    \or \mathsection
    \or \mathparagraph
    \or \|
    \or **
    \or \dagger\dagger
    \or \ddagger\ddagger
    \else
      \@ctrerr
    \fi
  }%
}
\makeatother

\makeatletter
\AtBeginDocument{\let\LS@rot\@undefined}
\makeatother

\usepackage{hyperref} 

\begin{document}

\preprint{APS/123-QED}

\title{Enriching molecular Raman spectroscopy with vibrational strong coupling}

\author{Matteo Castagnola$^{1}$}
\email{matcast@dtu.dk}

\author{Anne Todsen Hansen$^{2,3}$}

\author{Morten Hanefeld Dziegiel$^{2,4}$}

\author{Anders Kristensen$^{3}$}

\author{Søren Raza$^{1}$}

\author{Simone Latini$^{1}$}
\email{simola@dtu.dk}

\affiliation{$^{1}$Department of Physics, Technical University of Denmark, Fysikvej, 2800 Kongens Lyngby, Denmark}

\affiliation{$^{2}$Department of Clinical Immunology, Copenhagen University Hospital (Rigshospitalet), Blegdamsvej 9, 2100 København Ø, Denmark}

\affiliation{$^{3}$Department of Health Technology, Technical University of Denmark, Ørsteds Plads, Building 345C, 2800 Kongens Lyngby, Denmark}

\affiliation{$^{4}$Department of Clinical Medicine, University of Copenhagen, Blegdamsvej 3, 2200 København N, Denmark}

\begin{abstract}
Raman spectroscopy is widely used for molecular identification in biological samples, but spectral congestion often obscures key molecular fingerprints. 
Strategies to enrich vibrational spectra with additional controllable features are therefore desirable.
We theoretically show that vibrational strong coupling (VSC) can reshape Raman spectra by reorganizing vibrational energies and intensities.
Near-resonant coupling enables the resolution of quasi-degenerate vibrational modes and redistributes Raman activity among molecular vibrations, brightening otherwise Raman-inactive modes. 
Off-resonant coupling mediates effective interactions between vibrations, leading to tunable intensity reorganizations and spectral shifts via cavity-controlled vibrational mixing.
Using a microscopic theory of Raman scattering, we show that observing Raman signals from collective VSC polaritons requires appropriate geometries for the scattering setup, the sample, and the cavity environment, helping to explain why polaritonic Raman signatures have remained challenging to observe experimentally.

\end{abstract}

  \maketitle

\section*{Introduction}

Raman spectroscopy provides molecular fingerprints that are highly sensitive to the chemical structure. 
Its non-destructive and rapid nature makes it widely used for molecular identification in chemical, biological, and industrial settings. \cite{eissa2024perils, barth2009biological, kuhar2018challenges, frauendorfer2017industrial, zorzi2025flow} 
However, Raman spectra often suffer from significant peak overlap and inhomogeneous broadening, hindering spectral interpretation.\cite{guo2021chemometric, liu2026recent, samuel2021selecting}
In biological samples, fluctuating background signals further complicate the analysis. 
Molecular characterization thus often relies on spectral matching against reference libraries, but accuracy remains limited due to, e.g., sensitivity to differences in baseline and noise between the recorded and library spectra.\cite{samuel2021selecting, guo2021chemometric, liu2026recent, teran2025open, chen2023library, jensen2024label}
These limitations motivate strategies to enrich vibrational spectra with additional tunable features, providing auxiliary information for molecular discrimination and machine-learning-based identification methods.
Vibrational strong coupling (VSC) offers such an opportunity.

\begin{figure*}
    \centering
    \includegraphics[width=\linewidth]{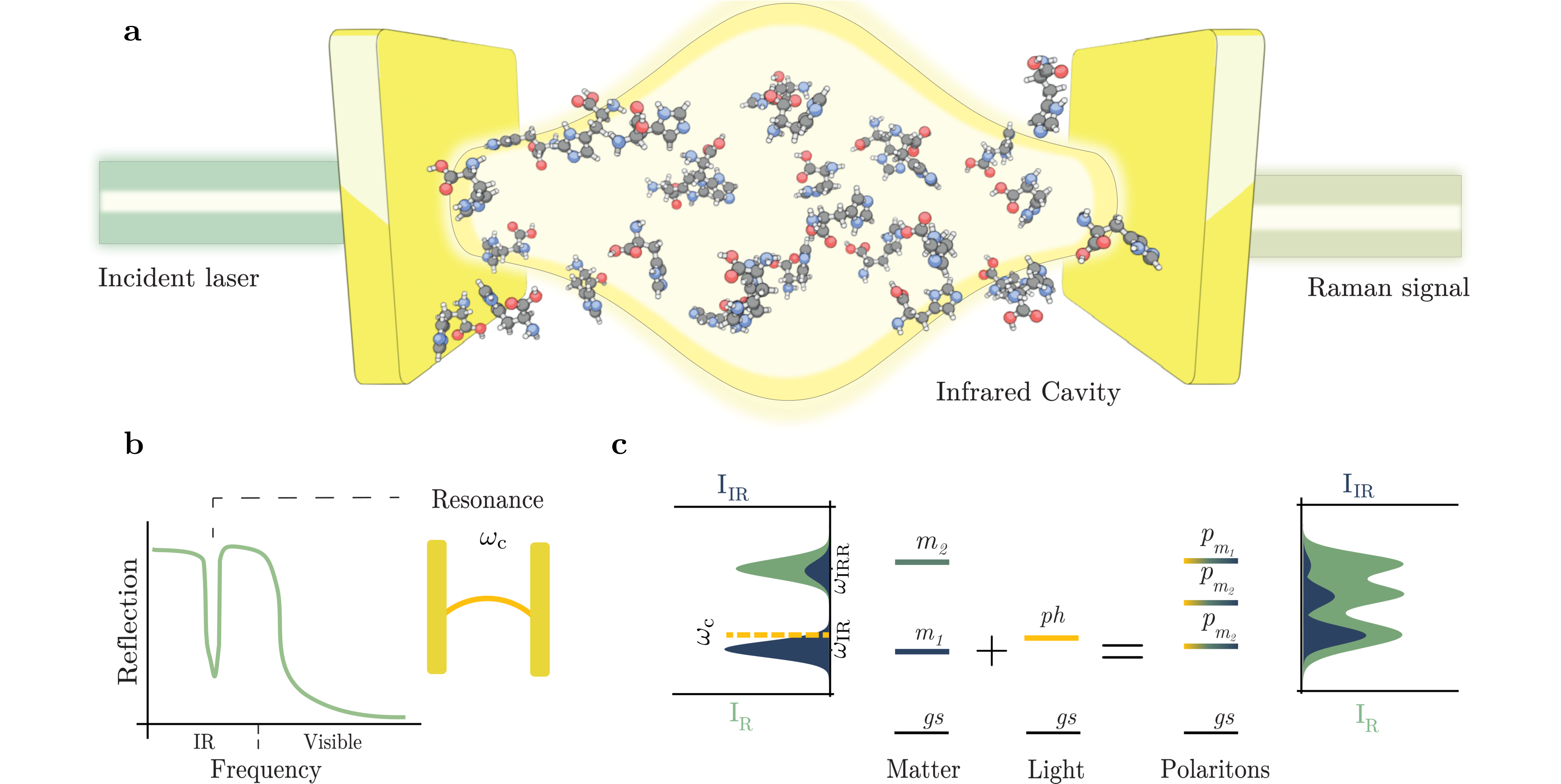}
    \caption{\textbf{Raman spectroscopy under vibrational strong coupling. }
    \textbf{a},
    Molecules confined within a Fabry-Pérot cavity collectively couple to an infrared cavity mode, forming vibrational polaritons. 
    The hybrid light-matter states are probed by an external visible laser, and the scattered Raman radiation is collected through the cavity mirrors.
    \textbf{b}, 
    Schematics of the assumed hot-mirror cavity reflectivity for Raman under VSC.
    The structure confines modes in the IR, while reflectivity drops in the visible, allowing the Raman laser to probe the molecules within.
    Such mirrors are formed from alternating metallic (e.g., Ag) and dielectric (e.g., TiO$_2$) layers, whose thicknesses set the visible and IR transmission.
    \textbf{c},
    Schematics of the interaction of IR-active ($\omega_\textrm{IR}$) and IR-Raman-active ($\omega_\textrm{IRR}$) molecular vibrations interacting with the cavity photon ($\omega_\textrm{c}$).
    The strong coupling leads to hybrid vibrational polaritons, which are IR-Raman-active and thus enable the detection of the Raman-dark mode $\omega_\textrm{IR}$. 
    }
    \label{fig: introfig}
\end{figure*}

Molecules confined within an optical resonator can collectively couple to its electromagnetic modes (Fig.~\ref{fig: introfig}a).
The interaction of molecular vibrations and cavity photons generates hybrid light-matter states, known as vibrational polaritons.
Their properties can be tuned by manipulating the strong coupling environment, such as the cavity frequency, mode structure, and molecular concentration, enabling controllable and non-invasive modification of molecular properties.\cite{wang2021roadmap, dunkelberger2022vibration, garcia2021manipulating}
We henceforth assume a hot-mirror cavity design, i.e., mirrors that strongly reflect IR radiation while transmitting visible light, so that a Raman laser can probe the molecules within (Fig.~\ref{fig: introfig}b). 
Such mirrors are routinely realized as thin-film stacks of silver and dielectric layers, as used in low-emissivity coatings.\cite{ding2017silver, hotmirror}
Polaritons can thus offer additional opportunities to manipulate Raman spectral information (Fig.~\ref{fig: introfig}c), giving optical access to a collective quantum state in which a vibrational excitation is coherently delocalized over many molecules.\cite{zheng2025quantum, zheng2023quantum, Vento2023MeasurementInduced}
Nevertheless, the experimental conditions under which polaritonic features become observable in Raman scattering remain unclear, and experimental observations have been mixed.
Initial reports suggested modifications of Raman spectra under VSC,\cite{shalabney2015enhanced, del2015signatures,strashko2016raman} whereas subsequent studies did not observe Raman-active polaritonic features.\cite{takele2021scouting, ahn2020raman, cohn2023spontaneous}
Recent theory suggests that Raman visibility depends on the sample geometry, with quasi-two-dimensional configurations providing favorable conditions for observing polaritonic Raman signals.\cite{dherbecourt2025spontaneous}
Clarifying the conditions under which VSC is observable in Raman spectra is essential for exploiting polaritonic spectroscopy in molecular sensing.
\\

In this work, we combine a microscopic theory of polaritonic Raman scattering and \textit{ab initio} simulations to show how VSC enriches Raman spectra and under which experimental conditions these features are observable.
Even though VSC creates coherent and collective light-matter states, Raman scattering can probe these states efficiently only when the incident and scattered fields interfere constructively across the molecular ensemble.
This phase-matching condition determines how the polaritonic signal scales with sample size and helps rationalize why Raman-active polaritons have remained difficult to observe.\cite{shalabney2015enhanced, del2015signatures,strashko2016raman, dherbecourt2025spontaneous, takele2021scouting, ahn2020raman, cohn2023spontaneous}
Our simulations demonstrate that light-matter hybridization can help resolve quasi-degenerate vibrational modes, activate Raman-inactive modes via intensity borrowing mechanisms, and induce tunable spectral reorganizations through cavity-mediated interactions.
These effects generate additional spectroscopic fingerprints that are inaccessible in conventional molecular Raman and IR spectroscopy.
We illustrate the mechanisms across several amino acids, chosen as chemically relevant systems for vibrational characterization in biological contexts.

\section*{Results}

\subsection*{Theoretical framework}

We consider a Fabry-Pérot cavity modeled as a single IR mode of frequency $\omega_\textrm{c}$ coupled to $N\gg 1$ molecules with $M$ vibrational modes of frequency $\omega_{m}$.
The system is described by the Tavis-Cummings Hamiltonian in the rotating-wave approximation
\begin{align}
    \hat H = &\omega_\textrm{c} \hat b ^\dagger \hat b + \sum_{i=1}^N \sum_{{m}=1}^M \omega_{m} \hat a^\dagger_{i,m} \hat a_{i,m}   \nonumber \\
    & + \sum_{i=1}^N \sum_{{m}=1}^M \bigg(g_{i,m} \hat b\; \hat a^\dagger_{i,m} +g_{i,m}^* \hat b ^\dagger \hat a_{i,m} \bigg), \label{eq: vibropol hamiltonian}
\end{align}
where $\hat b^\dagger$ creates a cavity photon, $\hat a^\dagger_{i,m}$ creates a vibrational excitation of mode ${m}$ in molecule $i$ located at position $\bm{R}_i$.
The coupling $g_{i,m} = \lambda \sqrt{\frac{\omega_\textrm{c}}{2}}  (\bm e_\textrm{c}\cdot \bm{\mu}_{i,m})e^{i \bm k_\textrm{c} \cdot \bm R_i}$ describes the interaction of mode $m$ of molecule $i$ with the cavity field (within the long-wavelength approximation), where $\bm k_\textrm{c}$ and $\bm e_\textrm{c}$ are the cavity mode wave vector and polarization, respectively, $\lambda$ parameterizes the cavity field amplitude, and $\bm{\mu}_{i,m}$ is the vibrational transition dipole moment.
Since the couplings are small compared to the vibrational frequencies (far from the ultrastrong coupling regime) and our focus is on single-excitation transitions rather than ground-state dressing, the rotating-wave approximation is here well justified.
\\

The Hamiltonian in Eq.~(\ref{eq: vibropol hamiltonian}) is conveniently transformed by introducing the collective bright modes $\hat A^\dagger_{m} = \frac{1}{ g_{m}}\sum_{i=1}^N {g_{i,m}} \,\hat a^\dagger_{i,m} $, where $ g_{m} = \sqrt{\sum_{i=1}^N |g_{i,m}|^2}$, and $M(N-1)$ (collective) dark states that do not couple to the cavity mode.
The bright mode is thus a coupling-weighted superposition of molecular vibrations with relative phases locked by the cavity field, and its interaction with the electromagnetic mode gives rise to the polaritonic states.
In the limit of approximately uniform cavity coupling, where all molecules experience nearly the same interaction $g_{i,m}\approx g_{0,m}$, the bright mode reduces to the familiar symmetric collective mode 
\begin{equation}
    \hat { A}^\dagger_{m} = \frac{1}{\sqrt{N}}\sum_{i=1}^N \hat a^\dagger_{i,m},  \label{eq: symmetric collective mode}
\end{equation}
with collective coupling $g_{m}=g_{0,m}\sqrt{N}$.
These collective states carry the entire IR oscillator strength, so that polaritonic states are readily observed in IR spectra.
Observing polaritonic signatures in Raman spectra depends on how efficiently the Raman process can access the bright mode.
\\

In Raman spectroscopy, molecules are probed by visible light, and an incident photon is inelastically scattered through a second-order process involving a virtual electronic excitation, creating a vibrational excitation.\cite{barron2009molecular, craig2012molecular}
Since the hot-mirror cavity confines IR but transmits visible light, we assume here that the mirrors do not fundamentally affect the Raman laser, so that the scattering process is treated perturbatively in terms of the interaction with the free-space optical fields (see Fig.~\ref{fig: introfig}b). 
The corresponding Raman interaction term is
\begin{align}
\hat V_\nu = A_\nu \sum_{i=1} 
\left( e^{i\bm k_\nu\cdot\bm R_i}\hat b_\nu
+
e^{-i\bm k_\nu\cdot\bm R_i}\hat b_\nu^\dagger \right)
(\bm e_\nu\cdot\hat {\bm\mu}^{el}_i),
\end{align}
where $\nu = \textrm{I}, \textrm{S}$ corresponds to the incident ($\textrm{I}$) and scattered ($\textrm{S}$) optical fields, $\hat b^\dagger_{\nu}$ creates a photon with wavevector $\bm{k}_{\nu}$ and polarization $\bm e_{\nu}$, $\hat {\bm\mu}^{el}_i$ is the (electronic) dipole operator of molecule $i$, and $A_{\nu}$ is the field normalization factor.\cite{craig2012molecular}
The Raman intensity $I_\textrm{R}$ is obtained from the squared second-order transition amplitude connecting the initial and final states via $\hat V_\textrm{I}$ and $\hat V_\textrm{S}$.
The resulting intensity must also include an orientational average over the random (isotropic) molecular orientations, which differs for the uniform and the non-uniform coupling due to the orientation-dependent term $\propto (\bm e_\textrm{c} \cdot \bm{\mu}_{i,m})$ (see the Supporting Information for a derivation of the Raman intensities).
\\

The next sections show how polariton formation enriches the Raman spectra, focusing on the case of the symmetric collective mode in Eq.~\ref{eq: symmetric collective mode}.
We then return to the experimental conditions under which the necessary coherent signal can be detected.

\subsection*{Resolving quasi-degenerate Raman features}
\begin{figure*}
    \centering
    \includegraphics[width=\linewidth]{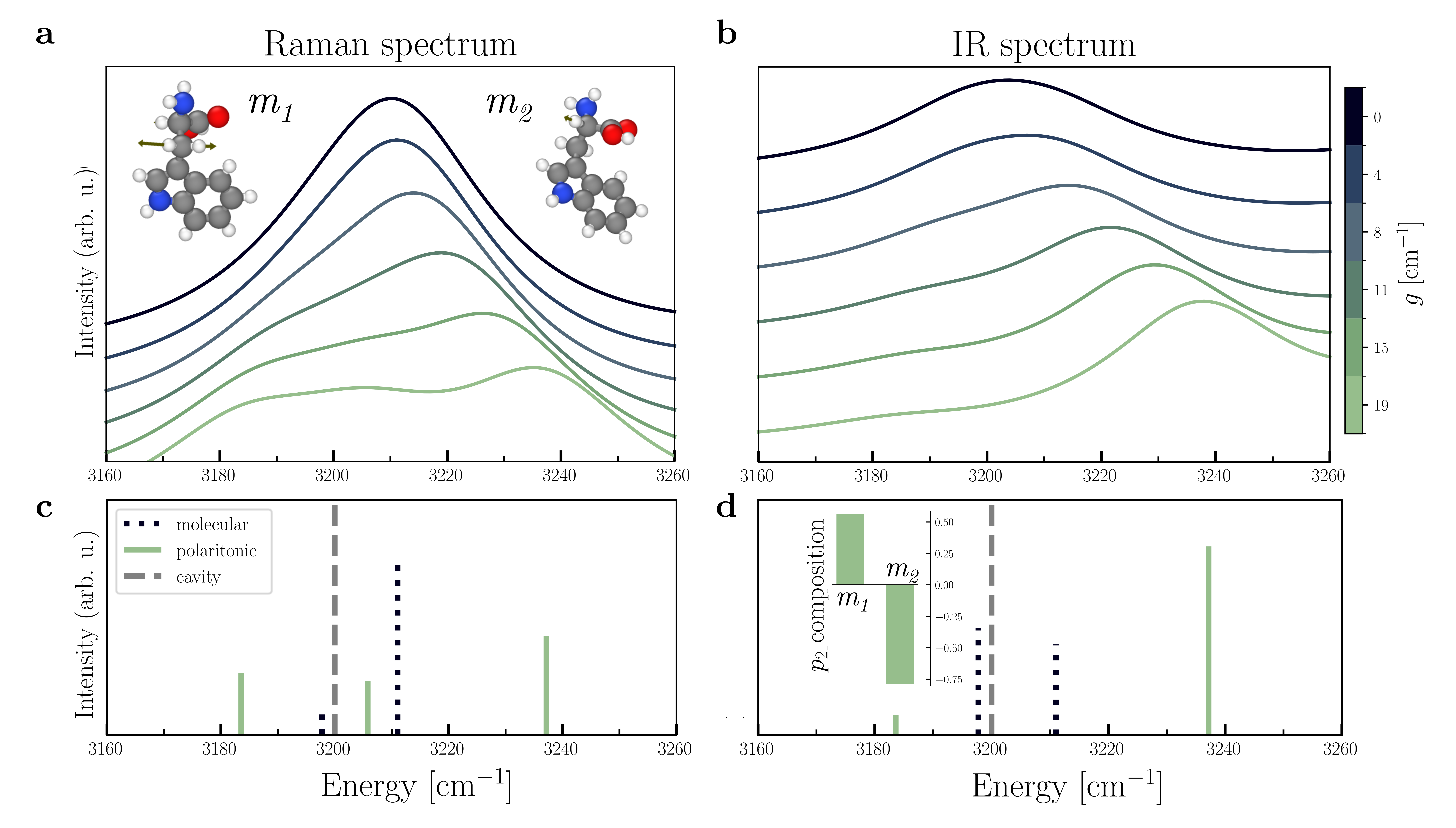}
    \caption{ \textbf{Resolving quasi-degenerate vibrations using Raman under VSC.}
    \textbf{a}, Raman spectra of tryptophan as the light-matter coupling strength increases from the molecular limit ($g=0$ cm$^{-1}$, blue) to the strong-coupling regime ($g=19$ cm$^{-1}$, green), for a cavity frequency $\omega_\textrm{c}=3200$ \si{\per\centi\meter}. 
    The light-matter interaction $g$ is quantified by the collective Tavis-Cummings coupling between the cavity mode and vibration $m_1$. 
    Two nearly degenerate molecular vibrations ($m_1$ and $m_2$, depicted on the right) hybridize with the cavity mode to form three polaritonic resonances ($p_1$, $p_2$, and $p_3$), which become spectrally resolved as the coupling increases. 
    \textbf{b}, 
    Corresponding IR spectra.
    Although the same molecular vibrations participate in the polaritons, the middle polariton $p_2$ carries only a weak IR oscillator strength and is not visible.
    \textbf{c}, 
    Line Raman spectrum for the largest coupling strength showing the cavity frequency (grey dashed line), together with the energies and Raman activities of the molecular vibrations (blue) and polaritonic states (green).
    \textbf{d}, Corresponding IR line spectrum.
    The bar plot insert shows the molecular composition of $p_2$. The opposite-sign contributions from $m_1$ and $m_2$ interfere destructively, suppressing the IR oscillator strength.
    }
    \label{fig: trypto 2}
\end{figure*}

Raman spectroscopy under VSC can reveal states that remain unresolved in conventional molecular Raman and IR spectroscopy.
To demonstrate this, we compute the Raman spectrum for tryptophan as the light-matter coupling strength is increased for a fixed cavity frequency of $\omega_\textrm{c}=3200$~\si{\per\centi\meter} (Fig.~\ref{fig: trypto 2}a).
In this energy range, tryptophan exhibits two nearly degenerate aliphatic C--H stretching modes, $m_1$ at $3197$~\si{\per\centi\meter} and $m_2$ at $3211$~\si{\per\centi\meter}.
In the uncoupled molecular spectra, these vibrations are not spectrally resolved because their energy separation is smaller than the linewidth, and the stronger Raman activity of $m_2$ dominates the signal.
As the light-matter coupling increases, the molecular vibrations hybridize with the cavity mode to form three polaritonic states, $p_1$, $p_2$, and $p_3$, with distinct energies and Raman intensities.
The energies and Raman activities of the individual polaritons show how the broad molecular feature ultimately resolves into three spectrally distinguishable resonances (Fig.~\ref{fig: trypto 2}c).
Interestingly, although the molecular IR absorption is determined by the same transition dipoles that govern the light-matter coupling, the IR spectrum does not permit identification of all polaritonic states (Fig.~\ref{fig: trypto 2}d).
The polariton $p_2$ remains essentially dark and is thus not identifiable (Fig.~\ref{fig: trypto 2}d). 
The origin of this dark state can be understood by examining the contribution of $m_1$ and $m_2$ (Tavis-Cummings coefficients) to the middle polariton (Fig.~\ref{fig: trypto 2}d).
The two vibrations contribute with comparable amplitudes but opposite signs, thereby interfering destructively and suppressing the IR oscillator strength.
This example demonstrates that VSC Raman spectroscopy can help resolve overlapping spectral features. 

\subsection*{Raman activity borrowing}
\begin{figure}
    \centering
    \includegraphics[width=\linewidth]{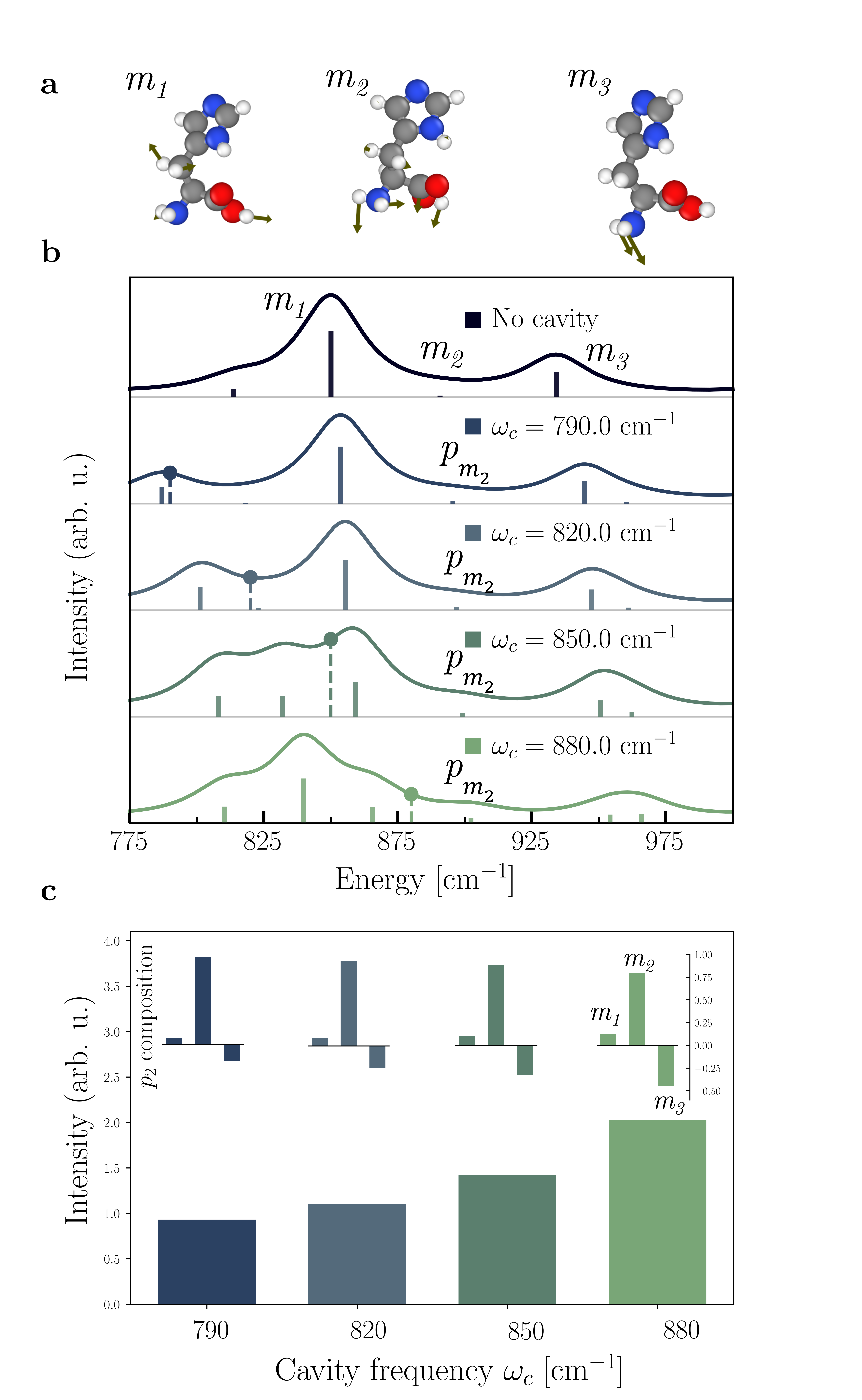}
    \caption{ \textbf{VSC activates Raman-inactive modes.}
    \textbf{a}, Visualization of the relevant molecular vibrations  of histidine ($m_1$, $m_2$, $m_3$).
    \textbf{b}, Raman spectra of histidine for different cavity frequencies $\omega_\textrm{c}$ (dashed lines).
    In the molecular limit (top), the vibration $m_2$ is nearly Raman inactive and masked by the stronger Raman-active modes $m_1$ and $m_3$. 
    The polaritonic state with dominant $m_2$ character $p_{m_2}$ gradually acquires Raman intensity via hybridization.
    \textbf{c}, Intensity of the predominantly-$m_2$ polariton as a function of cavity frequency. 
    Above, the molecular ($m_1$, $m_2$, $m_3$) components (Tavis-Cummings coefficients) of this state are shown for the corresponding cavity frequencies. 
    The increasing contribution from the Raman-active vibrations ($m_1$ and $m_3$) transfers Raman intensity to this otherwise nearly dark state.  
    }
    \label{fig: histidine}
\end{figure}

A second mechanism by which VSC enriches Raman spectra arises when the coupled vibrations have markedly different Raman strengths. 
In this case, cavity-induced hybridization redistributes the intensity among the polaritonic states, activating otherwise Raman-inactive vibrations.
In Fig.~\ref{fig: histidine}a, we show three IR-active vibrations ($m_1$, $m_2$, and $m_3$) of histidine that couple significantly to the cavity in the considered spectral region.
However, in the uncoupled Raman spectrum, only $m_1$ and $m_3$ contribute appreciably, while $m_2$ is nearly Raman-inactive (Fig.~\ref{fig: histidine}b).
Upon tuning the cavity frequency, the predominantly-$m_2$ polariton $p_{m_2}$ progressively gains Raman intensity.
The change is quantified in Fig.~\ref{fig: histidine}c, which displays the Raman intensity of $p_{m_2}$ together with its molecular composition.
As the cavity frequency is tuned, this state acquires increasing contributions from the Raman-active vibrations $m_1$ and $m_3$, as shown in the molecular components. 
The hybridization therefore transfers Raman intensity from the bright molecular vibrations to the otherwise nearly dark $m_2$ state, enhancing its visibility.

\subsection*{Off-resonant Raman intensity redistribution}
\begin{figure*}
    \centering
    \includegraphics[width=\linewidth]{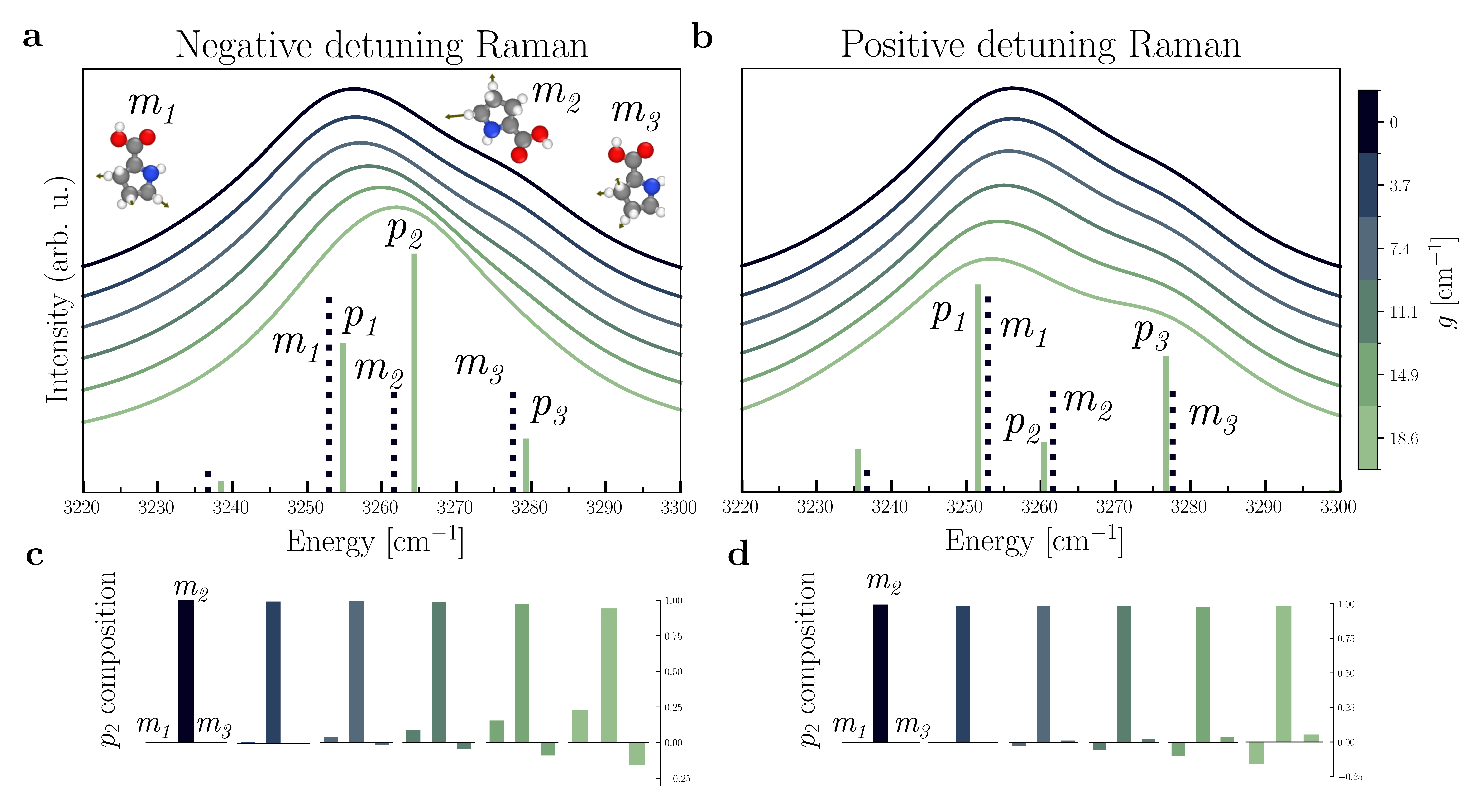}
    \caption{ \textbf{Off-resonance intensity restructuring mediated by virtual cavity photons.}
    \textbf{a}, Raman spectra of proline as the collective light-matter coupling $g$ between the cavity mode and vibration $m_1$ increases from the molecular limit ($g=0$ cm$^{-1}$, blue) to the strong-coupling regime ($g=18.6$ cm$^{-1}$, green), for a negatively-detuned cavity ($\omega_\textrm{c}=3100$ \si{\per\centi\meter}).
    The broadened Raman spectrum exhibits a blueshift due to energy shifts and intensity redistribution among the hybrid states.
    \textbf{b}, Corresponding results for a positively-detuned cavity ($\omega_\textrm{c}=3500$ \si{\per\centi\meter}).
    Here, the hybridization of $p_2$ differs qualitatively from that in panel \textbf{a}, leading to the development of a shoulder peak with modest energy shifts.
    \textbf{c}, Negative detuning molecular composition of the middle polariton $p_2$ as the coupling strength increases.
    \textbf{d}, Corresponding molecular components for positive detuning.
    }   
    \label{fig: proline 2}
\end{figure*}

We now show that VSC can also reshape Raman spectra for a strongly detuned cavity mode, where the sign of the detuning determines the qualitative spectral outcome.
Proline displays a broad Raman feature near $3250$~\si{\per\centi\meter} that changes markedly as both the cavity detuning $\Delta$ and the collective light-matter coupling $g$ are varied (Fig.~\ref{fig: proline 2}).
For negative detuning ($\Delta<0$), the Raman band shifts towards higher energies as the coupling strength increases (Fig.~\ref{fig: proline 2}a).
However, the underlying line spectrum shows that the polaritonic energies change modestly ($\sim 3$~\si{\per\centi\meter}) compared to the broad band ($\sim 6$~\si{\per\centi\meter}).
An inter-vibrational redistribution of Raman intensity contributes substantially to the observed blueshift, as evidenced by the line spectrum.
By contrast, for positive detunings ($\Delta>0$), the spectrum develops a pronounced shoulder within the broad spectral feature (Fig.~\ref{fig: proline 2}b).
\\

The observed spectral restructurings can be qualitatively understood in a perturbative framework.
For detunings much larger than the collective light-matter coupling, $|\Delta|\gg g$, virtual excitation of the cavity mode generates a (second-order) effective interaction between molecular vibrations $g_{\mathrm{eff}}\sim{g^2}/{\Delta}$.
The cavity thus mediates vibrational hybridization without being appreciably populated, with a mixing governed by the ratio $g_{\mathrm{eff}}/\delta\omega$, where $\delta\omega$ denotes the vibrational energy separation.
Therefore, molecular vibrations that are far from resonance with the cavity can hybridize appreciably if they lie sufficiently close in energy.
Moreover, $g_{\mathrm{eff}}$ inherits the sign of the cavity detuning $\Delta$.
The positive and negative detuning regimes thus lead to different molecular components in each polariton, as shown in Fig.~\ref{fig: proline 2}c,d.
The modified interference between the Raman scattering amplitudes produces the qualitatively different spectral reshapings (see the Supporting Information for the perturbative derivation).
These restructurings are reminiscent of the oscillator-strength redistribution in J- and H-aggregates, where the sign of the intermolecular dipole-dipole interaction is determined by the molecular packing geometry.\cite{hestand2018expanded, schembri2025supramolecular}
In the present case, the interaction is indirect, arising from a second-order virtual-photon process, and its sign can be reversibly modulated by adjusting the cavity detuning.

\subsection*{When are polaritonic Raman signals observable?}\label{sec: discussion}

Our results demonstrate that VSC can provide multiple enriching effects in Raman spectra, including the resolution of quasi-degenerate vibrations, borrowing of Raman activity, and off-resonant intensity restructurings. 
Nevertheless, polaritonic features in Raman spectra remain hard to measure.\cite{takele2021scouting, ahn2020raman, cohn2023spontaneous, dherbecourt2025spontaneous}
We now discuss why such measurements are challenging and estimate the experimental conditions under which VSC features can be observed in Raman spectra.
\\

The difficulty arises from a mismatch between the coherence and length scales of the vibrational polaritons and spontaneous Raman scattering.
Spontaneous Raman scattering ordinarily probes molecules as independent emitters via optical fields with shorter wavelengths (visible light, about half a micrometer).
The process is incoherent because each molecule imprints its own phase onto the scattered field, arising from its vibrational motion, the incident laser phase at its position, and the propagation distance to the detector.\cite{barron2009molecular}
By contrast, polaritons are composed of vibrational excitations that are coherent across the molecular ensemble (i.e., each molecule contributes with a definite amplitude and phase set by its cavity coupling) and delocalized over the IR length scale set by the Fabry-Pérot cavity (a few micrometers along the cavity axis).\\

The extended nature of polaritons thus raises the question of how the contributions of different molecules should be combined.
As the bright state involves a sum over all molecules, squaring the scattering amplitude generates terms involving the same molecule (local contribution) and terms involving two distinct molecules (nonlocal contribution). 
The former describes independent molecular scattering, while the latter arises from interference between molecular pairs.
For ordinary Raman, the nonlocal terms average to zero due to random molecular phases, leaving only the local contribution, which scales linearly with the number of molecules.
Here, because the collective state is normalized by $ 1/\sqrt{N}$, the one-molecule (incoherent) contribution is suppressed to $O(1)$.
The two-molecule contribution, by contrast, can recover the experimentally observable $O(N)$ scaling, provided that the scattering amplitudes from different molecules interfere constructively.
\\

Recording the coherent component of the Raman signal is not straightforward and imposes constraints on the experimental setup, helping to explain why polaritonic Raman signals remain elusive.\cite{shalabney2015enhanced, takele2021scouting, ahn2020raman, cohn2023spontaneous}
VSC provides extended coherence by locking the relative phases of the vibrational excitations in the collective bright state.
However, a coherent superposition further requires spatial phase matching of the incident and scattered optical (visible) fields across the ensemble, that is, the spatial phase $(\bm k_\textrm{I} - \bm k_\textrm{S})\cdot \bm R_i$ must vary little between molecules.
These additional conditions arise from averaging the Raman signal over the (random) orientations and positions of the molecules in the sample.
\\

The orientational average is more delicate than in ordinary spontaneous Raman scattering, because the cavity coupling is itself orientation-dependent $\propto(\bm e_\textrm{c}\cdot\bm\mu_{i})$.
As a result, the average differs between the uniform and non-uniform coupling regimes, and we report the full averaging procedure and the general case in the Supporting Information.
For the uniform coupling results shown here, the cavity polarization plays no role, and the scattered polarization follows that of the incident field, isolating the isotropic part of the Raman polarizability.
Recent effective theories indicate that such quasi-uniform approximation is a reasonable approximation.\cite{svendsen2025effective}
\\

\begin{figure}
    \centering
    \includegraphics[width=\linewidth]{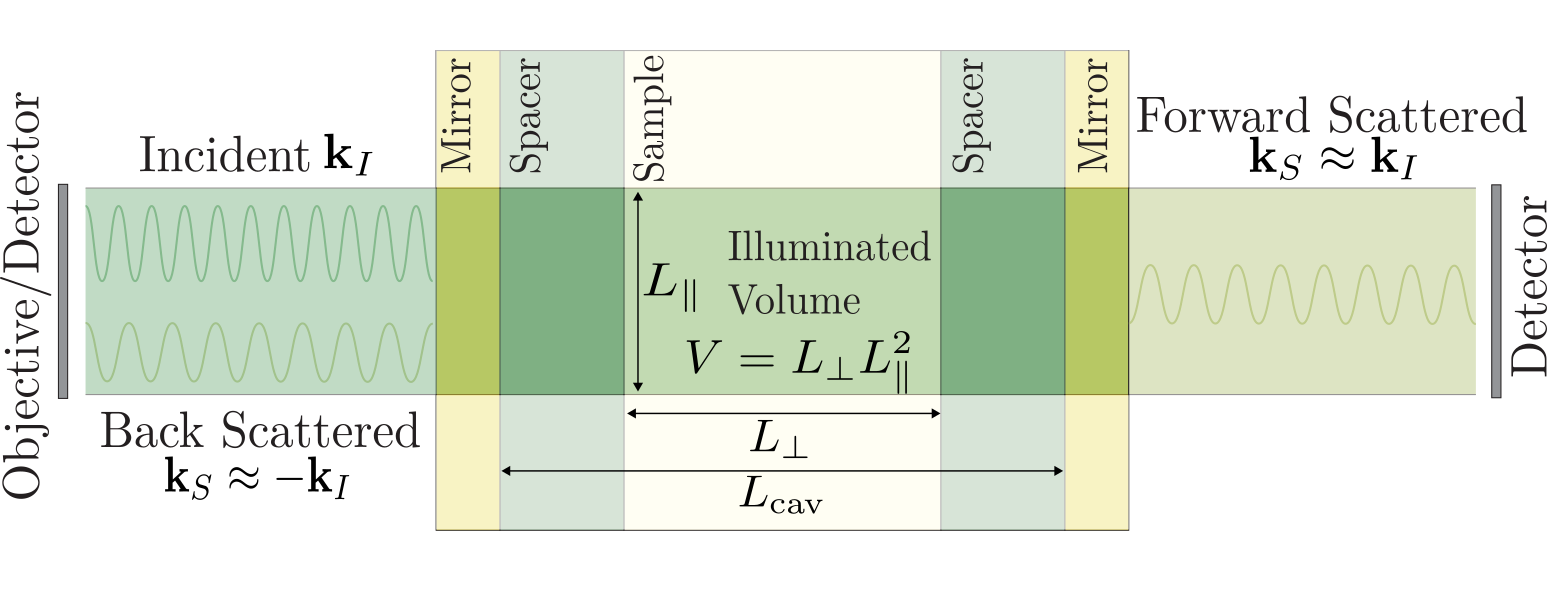}
    \caption{ \textbf{ 
    Scattering geometry for polaritonic Raman spectroscopy.}
    A molecular sample is confined between the mirrors of a Fabry-Pérot cavity by a dielectric spacer, which fixes the mirror separation $L_\textrm{cav}$ while allowing the molecular layer to occupy only part of the cavity volume ($L_\perp$ in the out-of-mirror-plane direction).
    Raman excitation is provided by a visible laser, incident perpendicularly to the cavity mirrors, with wave vector $\bm{k}_\textrm{I}$. The laser illuminates only part of the sample, corresponding to the volume $V=L_\perp L_\parallel^2$, where the in-plane dimension $L_\parallel$ is set by the laser beam width.
    The scattered light with wavevector $\bm k_\textrm{S}$ is collected at a detector (within a finite collection angle $\theta_0$, corresponding to numerical aperture $\text{NA} = \sin \theta_0$). 
}
    \label{fig: ramanscheme}
\end{figure}

Independent of the orientational average, the average over molecular positions imposes an additional geometric constraint.
Since the incident and scattered fields at the molecular site $\bm R_i$ carry position-dependent phases $e^{\pm i \bm k_{I,S}\cdot \bm R_i}$, each molecular contribution to the coherent amplitude bears a phase $e^{i(\bm k_\textrm{I}-\bm k_\textrm{S})\cdot \bm R_i}$.
A coherent sum thus requires that the accumulated mismatch over the illuminated volume remains small.
To quantify such a phase-matching condition, we consider the coherent Raman signal from a rectangular volume $V=L_\parallel^2L_\perp$, where $L_\parallel$ and $L_\perp$ denote the spatial extent of the sample illuminated by the incident laser (parallel and perpendicular to the mirrors, respectively) and thus contributing to the polaritonic Raman response (see Fig.~\ref{fig: ramanscheme}).
For the collective mode in Eq.~\ref{eq: symmetric collective mode}, we find that the coherent Raman intensity is proportional to a geometry-dependent factor
\begin{align}
    I_\textrm{R} \propto   \prod_{\alpha=x,y,z} \text{sinc}^2\bigg((\bm{k}_\textrm{I}-\bm{k}_\textrm{S} )_\alpha \frac{L_\alpha}{2}\bigg), \label{eq: intensity coherence}
\end{align}
where the sinc functions quantify the optical phase mismatch across $V$.
Eq.~(\ref{eq: intensity coherence}) shows that the Raman intensity signal is appreciable only when the optical phase varies little across the volume, thus quantifying the phase-matching condition that, together with the orientational requirements, governs the observability of the coherent polaritonic Raman signal.
Including the cavity-field envelope in the analysis contributes a further phase $e^{i \bm k_\textrm{c}\cdot \bm R_i}$, but it does not alter the derived scaling argument (see Supporting Information).
\\

Eq.~(\ref{eq: intensity coherence}) provides a criterion for estimating the relevant experimental conditions required to observe polaritonic Raman scattering.
This phase-matching condition is better interpreted in terms of the wavelength associated with the momentum transfer $\lambda_\alpha=\frac{2\pi}{\Delta k_\alpha}$, where $\Delta k_\alpha=|(\bm{k}_\textrm{I}-\bm{k}_\textrm{S} )_\alpha|$.
The coherent contribution remains close to its maximum provided that the  $L_\alpha$ is not much larger than $\lambda_\alpha$,
\begin{equation}
    \frac{L_\alpha}{\lambda_\alpha}\lesssim \frac{\Gamma}{\pi}, \label{eq: exp estimate}
\end{equation}
where $\Gamma\approx 1.39$ corresponds to the half-maximum of the sinc$^2(x)$ function.
Eq.~(\ref{eq: exp estimate}) compares the size of the illuminated ensemble $L_\alpha$ with the scale $\lambda_\alpha$ defined by the experimental setup, and thus restricts the system dimensions.
\\

In a back-scattering geometry, $\bm{k}_\textrm{I}-\bm{k}_\textrm{S}\approx 2\bm{k}_\textrm{I}$, so that the relevant wavelength is approximately half the incident wavelength $\lambda_\textrm{I}$.
For a visible laser ($\lambda_\textrm{I}\sim 500$ \si{\nano\meter}) traveling perpendicular to the mirrors, \autoref{eq: exp estimate} requires $L_\perp\lesssim 110$ \si{\nano\meter}.
This size is far smaller than the wavelength of a typical mid-infrared vibration, implying that the molecular ensemble must be tightly confined inside a Fabry-Pérot cavity using an appropriate spacer (see Fig.~\ref{fig: ramanscheme}). 
\\

In contrast, forward-scattering geometries are more favorable because $\bm{k}_\textrm{I}\approx\bm{k}_\textrm{S}$ in magnitude and direction.
The residual mismatch $\Delta\bm k$ is set by the small energy separation between the incident and scattered photons, defined by the vibrational excitation.
For a representative molecular vibration at $ 2000$ \si{\centi\per\meter}, corresponding to wavelength ${\lambda_{vib}}= 5$ \si{\micro\meter}, the phase-matching condition requires $L_\perp\lesssim 2.2$ \si{\micro\meter}, comparable to the thickness of a resonantly coupled Fabry-Pérot cavity ($L_{cav}\sim 2.5$ \si{\micro\meter}).
Forward-scattering geometries are therefore more compatible with the dimensions required for VSC.
\\

Eq.~(\ref{eq: intensity coherence}) must also be satisfied along the dimension parallel to the mirrors $L_\parallel$.
While we assume the incident laser to be perpendicular to the mirrors, the scattered light is collected over a finite angular range $\theta_0$ and thus with an in-plane component up to $|\bm k_\textrm{S}|\sin\theta_0\approx|\bm k_\textrm{I}|\sin\theta_0$.
Eq.~(\ref{eq: intensity coherence}) thus requires $\theta_0\lesssim \frac{2\Gamma}{|\bm k_\textrm{I}|L_\parallel}$, which imposes a constraint on the numerical aperture NA of the collection optics, $\text{NA}=\sin\theta_0$.
For a representative beam width $L_\parallel\sim 1$~\si{\micro\meter}, this gives $\theta_0\lesssim 13$°, which remains experimentally accessible.
In back-scattering experiments, the same objective is typically used both to focus the excitation beam and to collect the Raman signal (\autoref{fig: ramanscheme}). 
Therefore, back-scattering experiments require a compromise between beam focusing and in-plane phase matching, whereas this trade-off is lifted in forward-scattering geometries.
\\

Together, the phase-matching and orientational requirements are consistent with the difficulty of measuring polaritonic features in Raman spectra.\cite{takele2021scouting, ahn2020raman, cohn2023spontaneous}
Nevertheless, neither is a fundamental obstruction: quasi-two-dimensional samples and forward-scattering geometries with small numerical apertures offer concrete routes towards observing the coherent polaritonic contribution.

\section*{Discussion}

In this work, we showed that vibrational strong coupling enriches Raman spectra through multiple mechanisms, demonstrated across representative amino acids.
VSC can resolve quasi-degenerate vibrational modes, activate otherwise weak Raman features, and induce tunable off-resonant spectral reorganizations.
These spectroscopic fingerprints are absent from the uncoupled Raman and IR spectra and provide additional information for distinguishing chemically similar molecular systems.
Our microscopic theory shows that observing Raman-active polaritons requires constructive interference of the scattered optical fields across the molecular ensemble.
Such interference is possible because the cavity establishes a collective vibrational excitation with a well-defined phase across the molecular ensemble.
Observing polaritonic Raman signals additionally requires satisfying the optical phase-matching conditions.
Our analysis identifies forward-scattering geometries, quasi-two-dimensional samples,\cite{dherbecourt2025spontaneous} and small numerical apertures as promising experimental strategies.
These phase-matching requirements explain why Raman signatures of polaritons can remain elusive despite well-resolved polaritonic features in IR spectroscopy.
Under the derived conditions, Raman scattering offers a route to interrogating a collective quantum state that can be prepared and tuned via VSC, complementing the IR vibropolaritonic techniques proposed for molecular quantum sensing.\cite{zheng2025quantum, zheng2023quantum}

\section*{Methods}

Vibrational normal modes, transition dipoles, and Raman polarizabilities were obtained with the $e^T$ program,\cite{folkestad20262} an open-source electronic-structure program, and molecular modes are visualized using OVITO.\cite{stukowski2010visualization}
All calculations were carried out at the Hartree-Fock level of theory using the 6-31G* basis set, and initial molecular geometries (before optimization) are given in the Supporting Information.
Raman intensities were evaluated within the Placzek approximation and assuming uniform cavity coupling ($g_\textrm{i,m}\equiv g_{0,m}\quad\forall i$).
While HF is a level of theory with limitations for vibrational frequencies and Raman intensities,\cite{pople1993scaling, neugebauer2002coupled} our aim is to demonstrate that the three enrichment mechanisms identified here are physically accessible, not to reproduce quantitatively precise spectra for these specific amino acids.
For the polaritonic spectra, the Raman activity of each hybrid mode $p$ of energy $\omega_p$ was computed as
\begin{equation}
S_p=\left|\sum_m \frac{v_m^p}{\sqrt{2\omega_{m}}}
\left.\frac{\partial \alpha^e_{\alpha\alpha}}{\partial Q_{m}}\right|_{Q=Q_0}\right|^2,
\end{equation}
where $v_m^p$ is the coefficient of molecular vibration $m$ in the polariton state $p$, $\alpha^e_{\alpha\alpha}$ is the electronic polarizability tensor traced over Cartesian components, $Q_{m}$ is the normal mode displacement with frequency $\omega_{m}$, and $Q_0$ is the (ground state) equilibrium geometry.
The IR oscillator strengths $f_p =  \frac{2}{3}\omega_p D_p$ are obtained from the dipole strength $D_p$ 
\begin{equation}
    D_p = \left| \sum_m v_m^p \frac{1}{\sqrt{2\omega_{m}}}\left|\frac{\partial {\bm{\mu}^e}}{\partial \bm Q_{m}}\right|_{Q=Q_0} \right|^2,
\end{equation}
where $\bm{\mu}^e$ is the ground state electronic dipole.
The molecular spectra are computed using the same formulas, with the light-matter coupling strength set to zero.
The spectra were obtained by applying Lorentzian broadening $F(\omega)$ with linewidth $\gamma=15$ cm$^{-1}$, representative of typical molecular vibrational linewidths in condensed-phase systems, and normalized to unit maximum intensity
\begin{equation}
    F(\omega) = \frac{1}{\pi} \frac{\gamma}{(\omega-\omega_p)^2+\gamma^2}.
\end{equation}
The collective coupling strength was chosen such that the resulting Rabi splittings fell within the experimentally observed range of $30$-$100$ cm$^{-1}$.\cite{wang2021roadmap, dunkelberger2022vibration, garcia2021manipulating}
Further computational details and a thorough derivation of the scattering amplitudes, including the molecular orientational average, are presented in the Supporting Information.

\section*{Data availability}
All input and output files are available in the Zenodo repository \href{https://doi.org/10.5281/zenodo.20625724}{10.5281/zenodo.20625724}.

\section*{Acknowledgements}
S.L. acknowledges support from the Villum foundation grant No. 72146, the European Union Marie Sklodowska-Curie Doctoral Network SPARKLE grant No. 101169225. 
S.R. acknowledges support by the Novo Nordisk Foundation (NNF24OC0096142, NNF25OC0095542). 
All the authors acknowledge the Novo Nordisk Foundation grant No. NNF24OC0095967.



\bibliography{bib.bib}

@article{del2015signatures,
  title={Signatures of vibrational strong coupling in Raman scattering},
  author={del Pino, Javier and Feist, Johannes and Garcia-Vidal, FJ},
  journal={The Journal of Physical Chemistry C},
  volume={119},
  number={52},
  pages={29132--29137},
  year={2015},
  publisher={ACS Publications}
}

@article{strashko2016raman,
  title={Raman scattering with strongly coupled vibron-polaritons},
  author={Strashko, Artem and Keeling, Jonathan},
  journal={Physical Review A},
  volume={94},
  number={2},
  pages={023843},
  year={2016}
}

@article{dherbecourt2025spontaneous,
   title = {Spontaneous Raman Scattering under Vibrational Strong Coupling: The Critical Role of Polariton Spatial Mode Coherence},
  author = {Dherb\'ecourt, Maxime and Bellessa, Jo\"el and Symonds, Cl\'ementine and Weick, Guillaume and Hagenm\"uller, David},
  journal = {Phys. Rev. Lett.},
  volume = {137},
  issue = {11},
  pages = {116901},
  numpages = {7},
  year = {2026},
  month = {Sep},
  publisher = {American Physical Society}
}

@article{shalabney2015enhanced,
  title={Enhanced raman scattering from vibro-polariton hybrid states},
  author={Shalabney, Atef and George, Jino and Hiura, Hidefumi and Hutchison, James A and Genet, Cyriaque and Hellwig, Petra and Ebbesen, Thomas W},
  journal={Angewandte Chemie International Edition},
  volume={54},
  number={27},
  pages={7971--7975},
  year={2015},
  publisher={Wiley Online Library}
}

@article{takele2021scouting,
  title={Scouting for strong light--matter coupling signatures in Raman spectra},
  author={Takele, Wassie Mersha and Piatkowski, Lukasz and Wackenhut, Frank and Gawinkowski, Sylwester and Meixner, Alfred J and Waluk, Jacek},
  journal={Physical Chemistry Chemical Physics},
  volume={23},
  number={31},
  pages={16837--16846},
  year={2021},
  publisher={Royal Society of Chemistry}
}

@article{ahn2020raman,
  title={Raman scattering under strong vibration-cavity coupling},
  author={Ahn, Wonmi and Simpkins, BS},
  journal={The Journal of Physical Chemistry C},
  volume={125},
  number={1},
  pages={830--835},
  year={2020},
  publisher={ACS Publications}
}

@article{eissa2024perils,
  title={The perils of molecular interpretations from vibrational spectra of complex samples},
  author={Eissa, Tarek and Voronina, Liudmila and Huber, Marinus and Fleischmann, Frank and {\v{Z}}igman, Mihaela},
  journal={Angewandte Chemie International Edition},
  volume={63},
  number={50},
  pages={e202411596},
  year={2024},
  publisher={Wiley Online Library}
}

@book{barth2009biological,
  title={Biological and biomedical infrared spectroscopy},
  author={Barth, Andreas and Haris, Parvez I},
  volume={2},
  year={2009},
  publisher={IOS press}
}

@article{kuhar2018challenges,
  title={Challenges in application of Raman spectroscopy to biology and materials},
  author={Kuhar, Nikki and Sil, Sanchita and Verma, Taru and Umapathy, Siva},
  journal={RSC advances},
  volume={8},
  number={46},
  pages={25888--25908},
  year={2018},
  publisher={Royal Society of Chemistry}
}

@article{frauendorfer2017industrial,
  title={Industrial application of Raman spectroscopy for control and optimization of vinyl acetate resin polymerization},
  author={Frauendorfer, Eric and Hergeth, Wolf-Dieter},
  journal={Analytical and bioanalytical chemistry},
  volume={409},
  number={3},
  pages={631},
  year={2017},
  publisher={Springer Nature BV}
}

@article{folkestad20262,
  title={eT 2.0: An efficient open-source molecular electronic structure program},
  author={Folkestad, Sarai Dery and Kj{\o}nstad, Eirik F and Paul, Alexander C and Myhre, Rolf H and Alessandro, Riccardo and Angelico, Sara and Balbi, Alice and Barlini, Alberto and Bianchi, Andrea and Cappelli, Chiara and others},
  journal={The Journal of Chemical Physics},
  volume={164},
  number={13},
  year={2026},
  publisher={AIP Publishing}
}

@article{cohn2023spontaneous,
  title={Spontaneous Raman scattering from vibrational polaritons is obscured by reservoir states},
  author={Cohn, Bar and Filippov, Tikhon and Ber, Emanuel and Chuntonov, Lev},
  journal={The Journal of Chemical Physics},
  volume={159},
  number={10},
  year={2023},
  publisher={AIP Publishing}
}

@article{garcia2021manipulating,
  title={Manipulating matter by strong coupling to vacuum fields},
  author={Garcia-Vidal, Francisco J and Ciuti, Cristiano and Ebbesen, Thomas W},
  journal={Science},
  volume={373},
  number={6551},
  pages={eabd0336},
  year={2021},
  publisher={American Association for the Advancement of Science}
}

@article{dunkelberger2022vibration,
  title={Vibration-cavity polariton chemistry and dynamics},
  author={Dunkelberger, Adam D and Simpkins, Blake S and Vurgaftman, Igor and Owrutsky, Jeffrey C},
  journal={Annual review of physical chemistry},
  volume={73},
  pages={429--451},
  year={2022},
  publisher={Annual Reviews}
}

@article{wang2021roadmap,
  title={A roadmap toward the theory of vibrational polariton chemistry},
  author={Wang, Derek S and Yelin, Susanne F},
  journal={ACS Photonics},
  volume={8},
  number={10},
  pages={2818--2826},
  year={2021},
  publisher={ACS Publications}
}

@book{barron2009molecular,
  title={Molecular light scattering and optical activity},
  author={Barron, Laurence D},
  year={2009},
  publisher={Cambridge University Press}
}

@book{craig2012molecular,
  title={Molecular quantum electrodynamics},
  author={Craig, David Parker and Thirunamachandran, Thiru},
  year={2012},
  publisher={Courier Corporation}
}

@article{guo2021chemometric,
  title={Chemometric analysis in Raman spectroscopy from experimental design to machine learning--based modeling},
  author={Guo, Shuxia and Popp, J{\"u}rgen and Bocklitz, Thomas},
  journal={Nature protocols},
  volume={16},
  number={12},
  pages={5426--5459},
  year={2021},
  publisher={Nature Publishing Group UK London}
}

@article{liu2026recent,
  title={Recent Advances in Raman Spectral Classification with Machine Learning},
  author={Liu, Yonghao and Wu, Yizhan and Wang, Junjie and Qi, Jiantao and Zhou, Changjing and Xue, Yuhua},
  journal={Sensors},
  volume={26},
  number={1},
  pages={341},
  year={2026},
  publisher={MDPI}
}

@article{samuel2021selecting,
  title={On selecting a suitable spectral matching method for automated analytical applications of Raman spectroscopy},
  author={Samuel, Ashok Zachariah and Mukojima, Ryo and Horii, Shumpei and Ando, Masahiro and Egashira, Soshi and Nakashima, Takuji and Iwatsuki, Masato and Takeyama, Haruko},
  journal={ACS omega},
  volume={6},
  number={3},
  pages={2060--2065},
  year={2021},
  publisher={ACS Publications}
}

@article{teran2025open,
  title={Open Raman spectral library for biomolecule identification},
  author={Ter{\'a}n, Marcelo and Ruiz, Jos{\'e} Javier and Loza-Alvarez, Pablo and Masip, David and Merino, David},
  journal={Chemometrics and Intelligent Laboratory Systems},
  volume={264},
  pages={105476},
  year={2025},
  publisher={Elsevier}
}

@article{chen2023library,
  title={Library-Based Raman Spectral Identification Using Multi-Input Hybrid ResNet},
  author={Chen, Tiejun and Baek, Sung-June},
  journal={ACS omega},
  volume={8},
  number={40},
  pages={37482--37489},
  year={2023},
  publisher={ACS Publications}
}

@article{stukowski2010visualization,
  title={Visualization and analysis of atomistic simulation data with OVITO--the Open Visualization Tool},
  author={Stukowski, Alexander},
  journal={Modelling and simulation in materials science and engineering},
  volume={18},
  number={1},
  pages={015012},
  year={2010}
}

@article{hestand2018expanded,
  title={Expanded theory of H-and J-molecular aggregates: the effects of vibronic coupling and intermolecular charge transfer},
  author={Hestand, Nicholas J and Spano, Frank C},
  journal={Chemical reviews},
  volume={118},
  number={15},
  pages={7069--7163},
  year={2018},
  publisher={ACS Publications}
}

@article{schembri2025supramolecular,
  title={Supramolecular Engineering of Narrow Absorption Bands by Exciton Coupling in Pristine and Mixed Solid-State Dye Aggregates},
  author={Schembri, Tim and Albert, Julius and Hebling, Hendrik and Stepanenko, Vladimir and Anhalt, Olga and Shoyama, Kazutaka and Stolte, Matthias and W\"urthner, Frank},
  journal={ACS Central Science},
  volume={11},
  number={3},
  pages={452--464},
  year={2025},
  publisher={ACS Publications}
}

@article{zheng2025quantum,
  title={Quantum vibropolaritonic sensing},
  author={Zheng, Peng and Semancik, Steve and Barman, Ishan},
  journal={Science Advances},
  volume={11},
  number={33},
  pages={eady7670},
  year={2025},
  publisher={American Association for the Advancement of Science}
}

@article{zheng2023quantum,
  title={Quantum plexcitonic sensing},
  author={Zheng, Peng and Semancik, Steve and Barman, Ishan},
  journal={Nano Letters},
  volume={23},
  number={20},
  pages={9529--9537},
  year={2023},
  publisher={ACS Publications}
}

@article{zorzi2025flow,
  title={Flow cell for high throughput Raman spectroscopy of non-transparent solutions},
  author={Zorzi, Filippo and Jensen, Emil Alstrup and Serhatlioglu, Murat and Bonfadini, Silvio and Dziegiel, Morten Hanefeld and Criante, Luigino and Kristensen, Anders},
  journal={Lab on a Chip},
  volume={25},
  number={1},
  pages={69--78},
  year={2025},
  publisher={Royal Society of Chemistry}
}

@article{svendsen2025effective,
  title={Effective equilibrium theory of quantum light-matter interaction in cavities for extended systems and the long wavelength approximation},
  author={Svendsen, Mark Kamper and Ruggenthaler, Michael and H{\"u}bener, Hannes and Sch{\"a}fer, Christian and Eckstein, Martin and Rubio, Angel and Latini, Simone},
  journal={Communications Physics},
  volume={8},
  number={1},
  pages={425},
  year={2025},
  publisher={Nature Publishing Group UK London}
}

@article{jensen2024label,
  title={Label-Free Blood Typing by Raman Spectroscopy and Artificial Intelligence},
  author={Jensen, Emil Alstrup and Serhatlioglu, Murat and Uyanik, Cihan and Hansen, Anne Todsen and Puthusserypady, Sadasivan and Dziegiel, Morten Hanefeld and Kristensen, Anders},
  journal={Advanced Materials Technologies},
  volume={9},
  number={2},
  pages={2301462},
  year={2024},
  publisher={Wiley Online Library}
}

@article{pople1993scaling,
  title={Scaling factors for obtaining fundamental vibrational frequencies and zero-point energies from HF/6--31G* and MP2/6--31G* harmonic frequencies},
  author={Pople, John A and Scott, Anthony P and Wong, Ming Wah and Radom, Leo},
  journal={Israel Journal of Chemistry},
  volume={33},
  number={3},
  pages={345--350},
  year={1993},
  publisher={Wiley Online Library}
}

@article{neugebauer2002coupled,
  title={Coupled-cluster Raman intensities: Assessment and comparison with multiconfiguration and density functional methods},
  author={Neugebauer, Johannes and Reiher, Markus and Hess, Bernd A},
  journal={The Journal of chemical physics},
  volume={117},
  number={19},
  pages={8623--8633},
  year={2002},
  publisher={American Institute of Physics}
}

@article{ding2017silver,
  title={Silver-based low-emissivity coating technology for energy-saving window applications},
  author={Ding, Guowen and Clavero, C{\'e}sar},
  journal={World's largest Science, Technology \& Medicine Open Access book publisher},
  pages={409--431},
  year={2017}
}

@article{hotmirror,
  title={Comparison of the Optical Properties of Different Dielectric Materials (SnO2, ZnO, AZO, or SiAlNx) Used in Silver-Based Low-Emissivity Coatings},
  author={Cueva, Ana and Enrique, Carretero},
  journal={Coatings },
  pages={1709},
  volume={13},
  number={10},
  year={2023}
}

@article{Vento2023MeasurementInduced,
  author  = {Vento, Valeria and Tarrago Velez, Santiago and Pogrebna, Anna and Galland, Christophe},
  title   = {Measurement-induced collective vibrational quantum coherence under spontaneous Raman scattering in a liquid},
  journal = {Nature Communications},
  year    = {2023},
  volume  = {14},
  number  = {1},
  pages   = {2818},
}

\cleardoublepage

\pagestyle{empty}
\begin{widetext}


\section*{Supplementary material}

In section I, we review the theory of Raman scattering from an ensemble of randomly oriented molecules. The theory is extended in section II to the Raman scattering from molecular polaritons. The computational details are described in section III.

## I. RAMAN SCATTERING FROM A MOLECULE AND MOLECULAR ENSEMBLE

We consider Stokes Raman scattering, in which an incident photon is annihilated, and a scattered photon is created while the molecule is transferred from its vibrational ground state to an excited vibrational level. We will restrict our discussion to non-resonant Raman processes.

The Raman scattering from a molecule placed at position $\boldsymbol{R}$ is a second-order process.[1] We consider a molecule with a ground state $|g\rangle$ (electronic and vibrational) and several excited vibronic states, denoted collectively by $|s\rangle$. The incident (I) laser field has frequency $\omega_I$ and wave vector $\boldsymbol{k}_I$ with $n_I$ photons, and the molecule is initially in its ground state $|g\rangle$, with no photons of scattered (S) frequency $\omega_S$ and wave vector $\boldsymbol{k}_S$. The initial state of the system $|I\rangle$ is thus

$$|I\rangle = |g\rangle \otimes |n_I, 0_S\rangle \equiv |g; n_I, 0_S\rangle\,, \tag{1}$$

where in the light state $|n_I, 0_S\rangle$, the first entry refers to the incident and the last to the scattered field. The final state $|F\rangle$ of the system after the scattering occurred is

$$|F\rangle = |\nu\rangle \otimes |n_I - 1, 1_S\rangle \equiv |\nu; n_I - 1, 1_S\rangle\,, \tag{2}$$

with $|\nu\rangle$ a vibrationally excited state of the electronic ground-state. The light-matter interaction operators (in the dipole approximation) for the incident and scattered light are (in atomic units)

$$V_{I/S} = A_{I/S}(e^{i\boldsymbol{k}_{I/S}\cdot\boldsymbol{R}}\, b_{I/S} + e^{-i\boldsymbol{k}_{I/S}\cdot\boldsymbol{R}}\, b^{\dagger}_{I/S})(\boldsymbol{\epsilon}_{I/S}\cdot\boldsymbol{\mu}), \tag{3}$$

where $b^{\dagger}_{I/S}$ creates a photon of the incident or scattered waves, $\boldsymbol{\epsilon}_{I/S}$ is the light polarization, $\boldsymbol{\mu}$ is the molecular dipole operator, and $A_{I/S}$ is the field normalization factor.[2] The second-order transition amplitude of the scattering process thus reads

$$\mathcal{R} = \sum_s \left( \frac{\langle F|V_S|s; n_I - 1, 0_S\rangle \langle 0_S, n_I - 1; s|V_I|I\rangle}{E_g - E_s + \omega_I} + \frac{\langle F|V_I|s; n_I, 1_S\rangle \langle 1_S, n_I; s|V_S|I\rangle}{E_g - E_s - \omega_S} \right) \tag{4}$$

* matcast@dtu.dk

† simola@dtu.dk

where $E_g$ and $E_s$ are the molecular ground and intermediate state energies, and the sum occurs over all the possible intermediate molecular states. Explicitly using the form of the interaction operators, we obtain

$$\mathcal{R} = A_I A_S \sqrt{n_I} e^{i(\boldsymbol{k}_I - \boldsymbol{k}_S)\cdot \boldsymbol{R}} \sum_s \left( \frac{\langle \nu | \boldsymbol{\epsilon}_S \cdot \boldsymbol{\mu} | s \rangle \langle s | \boldsymbol{\epsilon}_I \cdot \boldsymbol{\mu} | g \rangle}{E_g - E_s + \omega_I} + \frac{\langle \nu | \boldsymbol{\epsilon}_I \cdot \boldsymbol{\mu} | s \rangle \langle s | \boldsymbol{\epsilon}_S \cdot \boldsymbol{\mu} | g \rangle}{E_g - E_s - \omega_S} \right). \quad (5)$$

The Raman scattering intensity is proportional to the square modulus of Equation 5, and the scattering rate $\mathrm{d}\Gamma$ per unit of solid angle $\mathrm{d}\Omega$ is obtained from the density of photon states $\rho$ in free space

$$\rho(|\boldsymbol{k}|) = \frac{|\boldsymbol{k}|^2 \mathcal{V}}{c(2\pi)^3} \mathrm{d}\Omega, \quad (6)$$

where $c$ is the speed of light and $\mathcal{V}$ the electromagnetic quantization volume.[2] We thus obtain

$$\mathrm{d}\Gamma = (A_I A_S)^2 n_I \frac{|\boldsymbol{k}_S|^2 \mathcal{V}}{c(2\pi)^3} \left| \sum_s \left( \frac{\langle \nu | \boldsymbol{\epsilon}_S \cdot \boldsymbol{\mu} | s \rangle \langle s | \boldsymbol{\epsilon}_I \cdot \boldsymbol{\mu} | g \rangle}{E_g - E_s + \omega_I} + \frac{\langle \nu | \boldsymbol{\epsilon}_I \cdot \boldsymbol{\mu} | s \rangle \langle s | \boldsymbol{\epsilon}_S \cdot \boldsymbol{\mu} | g \rangle}{E_g - E_s - \omega_S} \right) \right|^2 \mathrm{d}\Omega. \quad (7)$$

The molecular response is thus dictated by the (transition) polarizability

$$\alpha_{ij} = \sum_s \left( \frac{\langle \nu | \mu_i | s \rangle \langle s | \mu_j | g \rangle}{E_g - E_s + \omega_I} + \frac{\langle \nu | \mu_j | s \rangle \langle s | \mu_i | g \rangle}{E_g - E_s - \omega_S} \right), \quad (8)$$

where $\mu_i$ is the $i$-th component of the dipole operator. It is now convenient to introduce several approximations for computing this quantity. Energy conservation imposes $\omega_I - \omega_S = \omega_\nu$. Since vibrational frequencies are typically much smaller than optical frequencies, we approximate $\omega_I \approx \omega_S$ in the denominators. Then, in the Born-Oppenheimer approximation, each molecular state is partitioned into an electronic contribution and a vibrational one associated with each electronic state. We thus write

$$|s\rangle \equiv |s_e, \nu_{s_e}\rangle, \quad (9)$$

where the subscript $e$ refers to the electronic state, with associated vibration $\nu_{s_e}$. The purely vibrational contributions in Equation 8 are generally disregarded since the vibrational transition dipole of the ground state

$$\langle \nu'_{g_e} | \langle g_e | \boldsymbol{\mu} | g_e \rangle | \nu_{g_e} \rangle \quad (10)$$

is generally much smaller than the corresponding electronic transition dipoles

$$\langle \nu_{s_e}, s_e | \boldsymbol{\mu} | g_e, \nu_{g_e} \rangle, \quad (11)$$

and the incident and scattered light is highly detuned from vibrational transitions. Taking into account only electronically excited states, we further approximate the energy in the denominator by disregarding the vibrational contribution in the energy differences

$$E_g - E_s = E_{g_e} - E_{s_e} + E_{\nu_{ge}} - E_{\nu_{se}} \approx E_{g_e} - E_{s_e}, \tag{12}$$

leaving only the electronic excitation energy differences. We then partition the sum over the intermediate states $|s\rangle$ as a sum over the electronic $\sum_{s_e}$ and the corresponding vibrational $\sum_{\nu_{se}}$ states

$$\alpha_{ij} \approx \langle \nu'_{g_e}| \sum_{s_e} \left( \frac{\langle g_e|\mu_i|s_e\rangle \left( \sum_{\nu_{se}} |\nu_{s_e}\rangle \langle \nu_{s_e}| \right) \langle s_e|\mu_j|g_e\rangle}{E_{g_e} - E_{s_e} + \omega_I} + \frac{\langle g_e|\mu_j|s_e\rangle \left( \sum_{\nu_{se}} |\nu_{s_e}\rangle \langle \nu_{s_e}| \right) \langle s_e|\mu_i|g_e\rangle}{E_{g_e} - E_{s_e} - \omega_I} \right) |\nu_{g_e}\rangle \, , \tag{13}$$

where we moved the sum over the vibrational states in the numerator since we removed any explicit reference to the vibrational states in the denominator. Invoking the resolution of the identity on the vibrational levels, we obtain the vibrationally-averaged electronic polarizability

$$\alpha_{ij} \approx \langle \nu'_{g_e}| \sum_{s_e} \left( \frac{\langle g_e|\mu_i|s_e\rangle \langle s_e|\mu_j|g_e\rangle}{E_{g_e} - E_{s_e} + \omega_I} + \frac{\langle g_e|\mu_j|s_e\rangle \langle s_e|\mu_i|g_e\rangle}{E_{g_e} - E_{s_e} - \omega_I} \right) |\nu_{g_e}\rangle \equiv \langle \nu'_{g_e}|\alpha^e_{ij}|\nu_{g_e}\rangle \, , \tag{14}$$

where the electronic polarizability depends parametrically on the nuclear coordinates $\boldsymbol{Q}$ via the electronic energies $E_e \equiv E_e(\boldsymbol{Q})$ and transition moments $\langle s_e|\mu_i|g_e\rangle \equiv \langle s_e|\mu_i|g_e\rangle(\boldsymbol{Q})$ in the Born-Oppenheimer picture $\alpha^e \equiv \alpha^e(\boldsymbol{Q})$. We further assume the initial vibrational state is the vibrational ground state $|\nu_{g_e}\rangle \equiv |0_{g_e}\rangle$, and expand the electronic polarizability as a function of the nuclear coordinates $\boldsymbol{Q}$ around the minimum energy of the electronic state $\boldsymbol{Q} = \boldsymbol{Q}_0$

$$\alpha^e_{ij} \approx \alpha^{\boldsymbol{Q}_0}_{ij} + \left.\frac{\partial \alpha^e_{ij}}{\partial \boldsymbol{Q}}\right|_{\boldsymbol{Q}=\boldsymbol{Q}_0} (\boldsymbol{Q} - \boldsymbol{Q}_0) \tag{15}$$

and recover the standard molecular picture of Raman scattering as originating from the derivative of the electronic polarizability with the nuclear coordinates (the fixed-reference $\alpha^{\boldsymbol{Q}_0}_{ij}$ does not contribute to the transition expression due to orthogonality of the vibrational states). At this point, we need to take into account the random orientation of the molecule.[3] The Raman scattering intensity is proportional to the contraction

$$\alpha_{ij}\alpha_{kl}\epsilon^S_i\epsilon^I_j\epsilon^S_k\epsilon^I_l, \tag{16}$$

so we need to consider the isotropic average of $\alpha_{ij}\alpha_{kl}$. For simplicity, we assume the incident field to be polarized along $\hat{x}$, and we collect the scattered light independent of polarization (perpendicular to $\boldsymbol{k}_S$). That is, the total intensity is proportional to the sum

$$\sum_{\boldsymbol{\epsilon}_S \perp \boldsymbol{k}_S} \alpha_{ix}\alpha_{kx}\epsilon_i^S\epsilon_k^S = \alpha_{xx}\alpha_{xx} + \alpha_{yx}\alpha_{yx}, \tag{17}$$

where we conventionally assumed perpendicular scattering. If we introduce a molecular reference frame tied to the molecule, where the molecular indices are represented by Greek letters, the orientational average requires computing the following orientational averages (for an isotropic sample)

$$\langle l_{x\alpha}l_{x\beta}l_{x\gamma}l_{x\delta}\rangle = \frac{1}{15}(\delta_{\alpha\beta}\delta_{\gamma\delta} + \delta_{\alpha\gamma}\delta_{\beta\delta} + \delta_{\alpha\delta}\delta_{\gamma\beta}) \tag{18}$$

$$\langle l_{y\alpha}l_{x\beta}l_{y\gamma}l_{x\delta}\rangle = \frac{1}{30}(-\delta_{\alpha\beta}\delta_{\gamma\delta} + 4\delta_{\alpha\gamma}\delta_{\beta\delta} - \delta_{\alpha\delta}\delta_{\gamma\beta}) \tag{19}$$

where $l_{i\alpha}$ is the direction cosine between the direction $i$ in the lab frame and the direction $\alpha$ of the molecular axis. We assume the Einstein summation convention. The Raman scattering intensity $I_R$ in the solid angle d$\Omega$ is thus proportional to

$$I_R \propto \alpha_{\alpha\alpha}\alpha_{\beta\beta} + 6\alpha_{\alpha\beta}\alpha_{\alpha\beta} + \alpha_{\alpha\beta}\alpha_{\beta\alpha}. \tag{20}$$

Finally, we note that the collected Raman signal depends on the signal from all the $N \gg 1$ molecules in the sample. The orientational average takes into account the different orientations of the molecules. For the signal scaling with the number of molecules, we notice that Raman is an incoherent process, so the resulting scattering from the whole ensemble is simply $N$ times the single-molecule results, where the position of the molecule entering as the phase factor $e^{i(\boldsymbol{k}_I - \boldsymbol{k}_S)\cdot\boldsymbol{R}}$ does not affect the observed intensity. To account for the experimentally collected Raman signal, these expressions must be integrated over the solid angle subtended by the numerical aperture of the collection optics.

## II. RAMAN SCATTERING FROM POLARITONS

### A. Tavis-Cummings polaritonic picture

We describe the vibrational polaritons within the Tavis-Cummings model, neglecting vibrational disorder, modeling the cavity as a single effective optical mode, and restricting the

light-matter interaction to the rotating wave approximation. We consider $N \gg 1$ randomly oriented molecules with $M$ harmonic vibrations of frequency $\{\omega_m\}_{m=1}^{M}$, coupled to a cavity mode of frequency $\omega_c$, in the IR spectral region. In the rotating wave approximation, the Hamiltonian reads

$$H = \omega_c b^\dagger b + \sum_{i=1}^{N}\sum_{m=1}^{M} \omega_m a_{i,m}^\dagger a_{i,m} + \sum_{i=1}^{N}\sum_{m=1}^{M} \Big( g_{i,m} b\, a_{i,m}^\dagger + (g_{i,m})^* b^\dagger a_{i,m} \Big), \tag{21}$$

where $b^\dagger$ creates a cavity photon, $a_{i,m}^\dagger$ creates a vibrational excitation in mode $m$ for molecule $i$, and $g_{i,m}$ is the coupling of mode $m$ of molecule $i$ to the optical field. The light-matter coupling strength, described in the dipole approximation, reads

$$g_{i,m} = e^{i\boldsymbol{k}_c\cdot\boldsymbol{R}_i} \lambda \sqrt{\frac{\omega_c}{2}} (\boldsymbol{\epsilon}^{(c)} \cdot \boldsymbol{\mu}_{i,m}) \equiv (\boldsymbol{\epsilon}^{(c)} \cdot \boldsymbol{\xi}_{i,m}) e^{i\boldsymbol{k}_c\cdot\boldsymbol{R}_i}, \tag{22}$$

where $\boldsymbol{k}_c$ and $\boldsymbol{\epsilon}^{(c)}$ are the cavity mode wave vector and polarization, respectively, $\boldsymbol{R}_i$ is the position of the $i$-th molecule, and the vector $\boldsymbol{\xi}_{i,m} = \lambda\sqrt{\frac{\omega_c}{2}}\boldsymbol{\mu}_{i,m}$ encodes the light-matter coupling (without the cavity phase). In Equation 22, $\lambda$ parameterizes the cavity field amplitude (single-molecule light-matter coupling strength), and $\boldsymbol{\mu}_{i,m}$ is the transition dipole moment between the ground $|0_i\rangle$ and first-excited $|1_{i,m}\rangle = a_{i,m}^\dagger |0_i\rangle$ vibrational states of the electronic ground state of mode $m$ for molecule $i$

$$\boldsymbol{\mu}_{i,m} = \langle 1_{i,m}| \langle g_e|\boldsymbol{\mu}_i|g_e\rangle |0_{g_e}\rangle \approx \frac{1}{\sqrt{2\omega_m}} \frac{\partial \langle g_e|\boldsymbol{\mu}_i|g_e\rangle}{\partial \boldsymbol{Q}_m}. \tag{23}$$

In Equation 23, we introduced the normal mode displacement $\boldsymbol{Q}_m$ for the vibrational frequency $\omega_m$, recovering the well-known result that IR activity is governed by the derivative of the ground state electronic dipole moment.

To diagonalize Equation 21, we introduce the collective bright mode

$$A_m^\dagger = \frac{1}{g_m} \sum_{i=1}^{N} g_{i,m}\, a_{i,m}^\dagger, \tag{24}$$

where

$$g_m = \sqrt{\sum_{i=1}^{N} |g_{i,m}|^2} \tag{25}$$

is the root-mean-square collective coupling, which ensures $[A_m, A_{m'}^\dagger] = \delta_{mm'}$. In the limit of large $N$, we replace the sum in Equation 25 by $N$ times the orientational average of the light-matter coupling strength $g_{i,m}$, obtaining

$$g_m \approx |\lambda|\sqrt{N}\sqrt{\frac{\omega_c}{2}\frac{|\boldsymbol{\mu}_m|}{3}} \equiv \sqrt{N} g_0, \tag{26}$$

where $|\boldsymbol{\mu}_m|$ is the modulus of the transition dipole moment of mode $m$. Notice that the coupling strength is collective and depending on the microscopic (positive) quantity $g_0 = |\lambda|\sqrt{\frac{\omega_c}{2}}\frac{|\boldsymbol{\mu}_m|}{3}$ and scales are $\sqrt{N}$. If we assume all the molecules have an identical coupling to the cavity field, i.e., $g_{i,m} \equiv g_m^0$, the mode reduces to the familiar equal-weight superposition of in-phase molecular vibrations from the whole ensemble[4]

$$A_m^\dagger = \frac{1}{\sqrt{N}}\sum_{i=1}^{N} a_{i,m}^\dagger. \tag{27}$$

We also introduce $M \times (N-1)$ dark state (uncoupled) collective vibrational modes

$$D_{k,m} = \sum_{i=1}^{N} X_{ki,m}a_{i,m} \qquad k = 1, \ldots, (N-1) \tag{28}$$

where, for each vibrational mode $m$, the rows $(X_{kn,m})_{n=1}^{N}$ form an orthonormal set spanning the subspace orthogonal to the coupling vector $(g_{1,m}, g_{2,m}, \ldots, g_{N,m})$ to ensure a unitary transformation between the set of $(a_{i,m})$ operators to the set of $(A_m, D_{k,m})$ operators. In this way, the Hamiltonian in Equation 21 can be written as

$$H = \omega_c b^\dagger b + \sum_{m=1}^{M} \omega_m A_m^\dagger A_m + \sum_{m=1}^{M}\sum_{k=1}^{N-1} \omega_m (D_{k,m})^\dagger D_{k,m} + \sum_{m=1}^{M}(g_m\, bA_m^\dagger + g_m^*\, b^\dagger A_m). \tag{29}$$

The dark states are uncoupled from the cavity, and they are unmodified normal modes compared to the uncoupled ($\lambda = 0$) system. The bright mode interacts with the cavity field, giving rise to new hybrid normal modes, the polaritons. The effect of the cavity is thus to single out one mode, the bright collective mode, to form the polaritons, which will be shifted in energy compared to the reference (unmodified) dark state manifold. The Hamiltonian is diagonalized in the cavity and bright-state basis by solving the eigenvalues of the TC matrix

$$H = \begin{pmatrix} \omega_c & g_1 & g_2 & \cdots & g_M \\ g_1 & \omega_1 & 0 & \cdots & 0 \\ g_2 & 0 & \omega_2 & \cdots & 0 \\ \vdots & \vdots & \vdots & \ddots & \vdots \\ g_M & 0 & 0 & \cdots & \omega_M \end{pmatrix}. \tag{30}$$

The eigenvectors $\boldsymbol{v}^p$, $p = 1, \ldots, M+1$ of Equation 30 $H\boldsymbol{v}^p = E_p\boldsymbol{v}^p$ provide the expansion of the $p$-th polaritonic mode $\alpha_p^\dagger$ in terms of the collective vibrations and the photon mode

$$\alpha_p^\dagger = v_0^p b^\dagger + \sum_{m=1}^{M} v_m^p A_m^\dagger, \tag{31}$$

where the first component $v_0^p$ refers to the photon, and the remaining $M$ components $v_m^p$ $m = 1, \ldots, M$ to the collective molecular vibrations.

### *1. General properties of the TC Hamiltonian matrix*

The Cauchy interlacing theorem[5] ensures the eigenvalues $E_i$ of Equation 30 satisfy

$$E_1 \leq \omega_1 \leq E_2 \leq \cdots \leq \omega_M \leq E_{M+1}, \tag{32}$$

where we assumed the vibrational frequencies $\omega_m$ and the polaritonic energies $E_p$ are presented in ascending order. Assuming the photon component of the eigenvector $v_0^p$ is not zero, the eigenvector can be written, up to a normalization factor, as

$$\boldsymbol{v}^p \sim \begin{pmatrix} -1 \\ \frac{g_1}{\omega_1 - E_p} \\ \frac{g_2}{\omega_2 - E_p} \\ \vdots \\ \frac{g_M}{\omega_M - E_p} \end{pmatrix}, \tag{33}$$

which shows how different vibrations behave in- or out-of-phase with each other and with the photon component, depending on their relative energy with respect to the polaritonic state. The modes that are quasi-resonant with the cavity frequency $\omega_c$ strongly hybridize with the photons, leading to large energy splittings. In the limit of a single molecular vibration perfectly resonant with the cavity, the TC model leads to two perfectly hybridized polaritonic modes, the lower $\alpha_L^\dagger$ and upper polaritons $\alpha_U^\dagger$

$$\alpha_L^\dagger = \frac{b^\dagger - A^\dagger}{\sqrt{2}} \tag{34}$$

$$\alpha_U^\dagger = \frac{b^\dagger + A^\dagger}{\sqrt{2}} \tag{35}$$

separated in energy by the Rabi splitting

$$\Omega_R \sim 2\sqrt{N} g^0. \tag{36}$$

The off-resonant modes will mix with the cavity photons according to the detuning $\Delta$. A consequence of the coupling to a common optical mode is the introduction of an *effective* coupling between the vibrational modes; that is, many modes will contribute to the same

hybrid state. Such an indirect interaction can lead to significant restructuring of properties even in the off-resonant regime. We can rationalize the off-resonant mode mixing by means of perturbation theory, partitioning the Hamiltonian matrix in Equation 30 into two subspaces, $P$ and $Q$, which include a cluster of quasi-degenerate molecular vibrations and all the other states (including the photon), respectively

$$\begin{pmatrix} H_{QQ} & H_{QP} \\ H_{PQ} & H_{PP} \end{pmatrix} \begin{pmatrix} \Psi_Q \\ \Psi_P \end{pmatrix} = E \begin{pmatrix} \Psi_Q \\ \Psi_P \end{pmatrix}, \tag{37}$$

which we can formally rewrite as an equation only for the components in $P$

$$\left(H_{PQ}(H_{QQ} - E)^{-1} H_{QP} + H_{PP}\right) \Psi_P = E \Psi_P. \tag{38}$$

If we approximate the exact energy $E$ as an average energy $\bar{E}$ for the states in subspace $P$, we have derived an effective coupling between the molecular vibrations which scales as $g_{eff} \sim g^2/(\omega_c - \bar{E}) \sim g^2/\Delta$. The validity of such a perturbative approach is governed by the unitless parameter $g/\Delta$. If the coupling strength is comparable to the energy spreading $\delta\omega$ in the $P$ subspace, this effective interaction leads to the relevant vibrational mixing, even though the photon component remains small. This can induce a large restructuring of the molecular properties, even though the energies will not be much modified, consistent with the interlacing constraints imposed by the Cauchy theorem in Equation 32.

### 2. *IR absorption from molecular polaritons*

The IR absorption of polariton $p$ is proportional to the square of the projection of the transition dipole moment onto the incident light polarization, here assumed to be along the cavity polarization $\boldsymbol{\epsilon}^{(c)}$

$$D_p = \left| \langle 0, \tilde{G} | \Big( \sum_i \boldsymbol{\mu}_i \cdot \boldsymbol{\epsilon}^{(c)} \Big) \alpha_p^\dagger | \tilde{G}, 0 \rangle \right|^2, \tag{39}$$

where we have introduced the notation for the *collective* vibro-electronic ground state $|\tilde{G}\rangle$

$$|\tilde{G}\rangle = \bigotimes_{i=1}^{N} |g_i\rangle \otimes |0_i\rangle \,. \tag{40}$$

Expanding the polaritonic operator into the microscopic contributions, we obtain

$$D_p = \left| \sum_i \sum_m \frac{g_{i,m}}{\bar{g}_m} v_m^p (\boldsymbol{\mu}_{i,m}^* \cdot \boldsymbol{\epsilon}^{(c)}) \right|^2 = \sum_i \sum_{m,m'} v_m^p v_{m'}^p \frac{g_{i,m} g_{i,m'}^*}{\bar{g}_m \bar{g}_{m'}} (\boldsymbol{\mu}_{i,m}^* \cdot \boldsymbol{\epsilon}^{(c)}) (\boldsymbol{\mu}_{i,m'} \cdot \boldsymbol{\epsilon}^{(c)})$$
$$+ \sum_{i \neq j} \sum_{m,m'} v_m^p v_{m'}^p \frac{g_{i,m} g_{j,m'}^*}{\bar{g}_m \bar{g}_{m'}} (\boldsymbol{\mu}_{i,m}^* \cdot \boldsymbol{\epsilon}^{(c)}) (\boldsymbol{\mu}_{j,m'} \cdot \boldsymbol{\epsilon}^{(c)}) \qquad (41)$$

Assuming identical molecules with homogeneous cavity coupling and an isotropic orientational distribution, we obtain

$$D_p \approx N \left| \sum_m v_m^p |\boldsymbol{\mu}_m| \right|^2 . \qquad (42)$$

### B. Raman scattering from the collective bright state

The observability of the Raman signal of polaritons depends on the Raman activity of the collective bright mode in Equation 24 or Equation 27. For the polariton mode in Equation 31, the corresponding Raman scattering amplitude reads

$$\mathcal{R} = \sum_s \left( \frac{\langle F|V_S|s; n_I - 1, 0_S\rangle \langle 0_S, n_I - 1; s|V_I|I\rangle}{E_I - E_s + \omega_I} + \frac{\langle F|V_I|s; n_I, 1_S\rangle \langle 1_S, n_I; s|V_S|I\rangle}{E_I - E_s - \omega_S} \right) \qquad (43)$$

where the initial state $|I\rangle$ and final state $|F\rangle$ read

$$|I\rangle = \bigotimes_{i=1}^N \left( |g_i\rangle \otimes_{m=0}^M |0_i^m\rangle \right) \otimes |n_I, 0_S\rangle \equiv |G, 0^G; n_I, 0_S\rangle \qquad (44)$$
$$|F\rangle = \alpha_p^\dagger \bigotimes_{i=1}^N \left( |g_i\rangle \ \otimes_{m=0}^M |0_i^m\rangle \right) \otimes |(n-1)_I, 1_S\rangle \equiv |G, 1_p^G; n_I - 1, 1_S\rangle \, . \qquad (45)$$

Here, we have introduced the notation for the *collective* electronic ground state $|G\rangle$, which is the tensor product of the electronic ground state $|g_i\rangle$ of each molecule $i$

$$|G\rangle = \bigotimes_{i=1}^N |g_i\rangle \, , \qquad (46)$$

and the collective vibrational ground state for each vibrational mode *and* the cavity mode, indexed as mode $m = 0$

$$|0^G\rangle = \bigotimes_{i=1}^N \otimes_{m=0}^M |0_{i,m}\rangle = |0\rangle \otimes \bigotimes_{i=1}^N \otimes_{m=1}^M |0_{i,m}\rangle \, , \qquad (47)$$

where the first ket refers to the cavity photon vacuum. The excited polaritonic (final) mode is

$$|1_p^G\rangle = \alpha_p^\dagger |0^G\rangle . \tag{48}$$

In Equation 43, the intermediate states $|s\rangle$ are vibronic excitations from all the molecules in the system (dressed by the cavity field). If we assume that the cavity does not confine optical modes in the visible region, which is reasonable for IR cavities, the interaction operators for the incident and scattered fields are taken to be identical to the free-space ones

$$V_{I/S} = A_{I/S} \sum_{i=1}^{N_I} (e^{i\boldsymbol{k}_{I/S}\cdot\boldsymbol{R}_i}\ b_{I/S} + e^{-i\boldsymbol{k}_{I/S}\cdot\boldsymbol{R}_i}\ b_{I/S}^\dagger)(\boldsymbol{\epsilon}_{I/S}\cdot\boldsymbol{\mu}_i), \tag{49}$$

where the sum runs over the $N_I$ molecules in the cavity illuminated by the external probe. In general, $N_I$ need not coincide with the number of molecules coherently coupled to the cavity mode $N_I \neq N$. Since the final polaritonic state is a superposition of collective vibrational excitations weighted by $v_m^p$, the Raman amplitude will inherit the same decomposition.

We now need to characterize the intermediate states $|s\rangle$ to compute the transition amplitude. The interaction operators excite and de-excite at most one molecule in the sample, so $|s\rangle$ will be restricted to the first excitation manifold. In particular, $V_I$ and $V_S$ must sequentially excite and de-excite the same molecule (processes in which different molecules are involved are beyond second-order). We can then make the same assumptions as for molecular Raman scattering, working in the Born-Oppenheimer approximation, thus assuming that $|s\rangle$ corresponds to an electronic excitation, and disregarding the differences between $\omega_I$ and $\omega_S$ and the vibrational energy in the denominators. We notice that the different electronic excitations of distinct molecules, $|s_e^i\rangle$ and $|s_e^j\rangle$ for $j \neq i$, are indirectly interacting via the IR cavity mode. Nevertheless, the cavity energy is far detuned from the electronic excitations, and the interaction is therefore second order. We thus choose to ignore such interaction, so that we can consider the intermediate states $|s_e\rangle$ as electronic excitations localized on the different molecules in the cavity. The Born-Oppenheimer approximation thus partitions the electronic excitation and the cavity-vibrational excitations, which we remove invoking the resolution of the identity as for Equation 13. The Raman amplitude thus becomes

$$\mathcal{R} \approx \sum_{m=1}^{M} v_m^p A_I A_S \sqrt{n_I} \sum_{i=1}^{N_I} e^{i(\boldsymbol{k}_I - \boldsymbol{k}_S)\cdot\boldsymbol{R}_i} \sum_{s_e^i} \Bigg( \frac{\langle 0^G, G| A_m(\boldsymbol{\epsilon}_S\cdot\boldsymbol{\mu}_i)|s_e^i\rangle\, \langle s_e^i|\boldsymbol{\epsilon}_I\cdot\boldsymbol{\mu}_i|G, 0^G\rangle}{E_{g_e} - E_{s_e} + \omega_I} + \frac{\langle 0^G, G| A_m(\boldsymbol{\epsilon}_I\cdot\boldsymbol{\mu}_i)|s_e^i\rangle\, \langle s_e^i|\boldsymbol{\epsilon}_S\cdot\boldsymbol{\mu}_i|G, 0^G\rangle}{E_{g_e} - E_{s_e} - \omega_I} \Bigg). \tag{50}$$

The molecular operator $\boldsymbol{\epsilon}_{I/S} \cdot \boldsymbol{\mu}_i$ then selects in the collective mode $A_m$ of Equation 24 the contribution of molecule $i$, yielding

$$\mathcal{R} \approx \sum_{m=1}^{M} v_m^p A_I A_S \sqrt{n_I} \sum_{i=1}^{N_I} e^{i(\boldsymbol{k}_I - \boldsymbol{k}_S)\cdot \boldsymbol{R}_i} \frac{(g_{i,m})^*}{\bar{g}_m} \sum_{s_e^i} \Bigg( \frac{\langle 0^G, G| a_{i,m} (\boldsymbol{\epsilon}_S \cdot \boldsymbol{\mu}_i) |s_e^i\rangle \langle s_e^i | \boldsymbol{\epsilon}_I \cdot \boldsymbol{\mu}_i | G, 0^G\rangle}{E_{g_e} - E_{s_e} + \omega_I} + \frac{\langle 0^G, G| a_{i,m} (\boldsymbol{\epsilon}_I \cdot \boldsymbol{\mu}_i) |s_e^i\rangle \langle s_e^i | \boldsymbol{\epsilon}_S \cdot \boldsymbol{\mu}_i | G, 0^G\rangle}{E_{g_e} - E_{s_e} - \omega_I} \Bigg) = \tag{51}$$

$$\sum_{m=1}^{M} v_m^p A_I A_S \sqrt{n_I} \sum_{i=1}^{N_I} e^{i(\boldsymbol{k}_I - \boldsymbol{k}_S)\cdot \boldsymbol{R}_i} \frac{(g_{i,m})^*}{\bar{g}_m} \langle 1_{i,m}| \sum_{s_e^i} \Bigg( \frac{\langle g_e^i| (\boldsymbol{\epsilon}_S \cdot \boldsymbol{\mu}_i) |s_e^i\rangle \langle s_e^i | \boldsymbol{\epsilon}_I \cdot \boldsymbol{\mu}_i | g_e^i\rangle}{E_{g_e} - E_{s_e} + \omega_I} + \frac{\langle g_e^i| (\boldsymbol{\epsilon}_I \cdot \boldsymbol{\mu}_i) |s_e^i\rangle \langle s_e^i | \boldsymbol{\epsilon}_S \cdot \boldsymbol{\mu}_i | g_e^i\rangle}{E_{g_e} - E_{s_e} - \omega_I} \Bigg) |0_{i,m}\rangle = \tag{52}$$

$$\sum_{m=1}^{M} v_m^p A_I A_S \sqrt{n_I} \sum_{i=1}^{N_I} e^{i(\boldsymbol{k}_I - \boldsymbol{k}_S)\cdot \boldsymbol{R}_i} \frac{(g_{i,m})^*}{\bar{g}_m} \langle 1_{i,m}| \, \alpha_{pq}^i \, |0_{i,m}\rangle \, \epsilon_p^S \epsilon_q^I, \tag{53}$$

that is, the Raman amplitude is given by a superposition of the contribution from the scattering of each molecule via their (local) transition polarizability $\alpha_{pq}^i$, weighted by a phase factor originating from the incident and scattered field $e^{i(\boldsymbol{k}_I - \boldsymbol{k}_S)\cdot \boldsymbol{R}_i}$, as well as the cavity contribution $g_{i,m}/\bar{g}_m$.

#### 1. *The problem of coherence*

The Raman intensity is thus proportional to the modulus squared of the amplitude scaled by the density of final scattered photon states

$$I_R \propto (A_I A_S)^2 n_I \frac{|\boldsymbol{k}_S|^2 \mathcal{V}}{c(2\pi)^3} \mathrm{d}\Omega \sum_{m,m'=1}^{M} v_m^p (v_{m'}^p)^* \sum_{i,j=1}^{N_I} e^{i(\boldsymbol{k}_I - \boldsymbol{k}_S)\cdot(\boldsymbol{R}_i - \boldsymbol{R}_j)} \frac{g_{j,m'} (g_{i,m})^*}{\bar{g}_m \bar{g}_{m'}} \langle 1_{i,m}| \, \alpha_{pq}^i \, |0_{i,m}\rangle \langle 1_{j,m'}| \, \alpha_{kl}^j \, |0_{j,m'}\rangle^* \epsilon_k^S \epsilon_l^I \epsilon_p^S \epsilon_q^I \tag{54}$$

in the small solid angle $\mathrm{d}\Omega$, and $\mathcal{V}$ is the electromagnetic quantization volume. We can thus substitute the molecular quantities by their (orientational) average value $\langle \ldots \rangle_O$, leading us

to (assuming molecular orientations and positions are uncorrelated)

$$I_R \propto (A_I A_S)^2 n_I \frac{|\boldsymbol{k}|^2 \mathcal{V}}{c(2\pi)^3} \mathrm{d}\Omega \sum_{m,m'=1}^{M} v_m^p (v_{m'}^p)^* \left\langle \frac{\tilde{g}_{j,m'}(\tilde{g}_{i,m})^*}{\bar{g}_m \bar{g}_{m'}} \langle 1_{i,m} | \alpha_{pq}^i | 0_{i,m} \rangle \langle 1_{j,m'} | \alpha_{kl}^j | 0_{j,m'} \rangle^* \epsilon_k^S \epsilon_l^I \epsilon_p^S \epsilon_q^I \right\rangle_O \times \sum_{i,j=1}^{N_I} e^{i(\boldsymbol{k}_I - \boldsymbol{k}_S)\cdot(\boldsymbol{R}_i - \boldsymbol{R}_j)} \phi_{mm'}(i,j), \tag{55}$$

where in the light-matter couplings we separated the phase factor coming from the molecular positions $g_{i,m} = \tilde{g}_{i,m}\phi_m(i)$ and $\phi_{mm'}(i,j) = \phi_m^*(i)\phi_{m'}(j)$. There are two important points to highlight in Equation 55. First, the product of the root-mean-square collective couplings $g_m g_{m'}$ scales as $N$ since $g_m \sim \sqrt{N} g_0$. Second, independently of the orientational averaging problem, the expression carries a structure factor (including the $1/N$ from the collective coupling strengths)

$$\frac{1}{N}\left|\sum_{i=1}^{N_I} e^{i(\boldsymbol{k}_I - \boldsymbol{k}_S)\cdot\boldsymbol{R}_i} \phi_m(i)\right|^2, \tag{56}$$

which depend on the molecular positions as they are probed by the incident, scattering, and cavity field. This structure factor carries a relevant phase-matching condition that strongly influences the observability of the polaritonic Raman signal. We note that the incoherent part $i = j$, which scales as $N$, is suppressed by the $1/N$ prefactor. Therefore, the incoherent contribution does not scale with the number of molecules $O(1)$, making it strongly suppressed and likely difficult to observe experimentally. Therefore, observation of Raman scattering from molecular polaritons requires coherence ($N_I^2$ scaling, which is reduced to $N_I^2/N$ by the prefactor). For a large number of molecules, the sum can be converted to an integral (assuming uniform molecular density $\rho$)

$$\frac{1}{N}\sum_{i,j=1}^{N_I} e^{i(\boldsymbol{k}_I - \boldsymbol{k}_S)\cdot(\boldsymbol{R}_i - \boldsymbol{R}_j)} \phi_{mm'}(i,j) \to \frac{\rho^2}{N} \iint \mathrm{d}^3 r_1 \mathrm{d}^3 r_2 e^{i(\boldsymbol{k}_I - \boldsymbol{k}_S)\cdot(\boldsymbol{r}_1 - \boldsymbol{r}_2)} \phi_{mm'}(\boldsymbol{r}_1, \boldsymbol{r}_2). \tag{57}$$

As a first approximation, we assume the coupling to the cavity to be spatially uniform, i.e., $\phi_m(\boldsymbol{r}) = 1$. For a parallelepiped illuminated volume $V = L_x L_y L_z$

$$\frac{\rho^2}{N} \iint \mathrm{d}^3 r_1 \mathrm{d}^3 r_2 e^{i(\boldsymbol{k}_I - \boldsymbol{k}_S)\cdot(\boldsymbol{r}_1 - \boldsymbol{r}_2)} = \frac{\rho^2}{N} V^2 \prod_{\alpha=x,y,z} \mathrm{sinc}^2\left((\boldsymbol{k}_I - \boldsymbol{k}_S)_\alpha \frac{L_\alpha}{2}\right) = \frac{N_I^2}{N} \prod_{\alpha=x,y,z} \mathrm{sinc}^2\left((\boldsymbol{k}_I - \boldsymbol{k}_S)_\alpha \frac{L_\alpha}{2}\right), \tag{58}$$

which shows that the collective Raman signal scales as $N$ only when the phase mismatch $\boldsymbol{k}_I - \boldsymbol{k}_S$ is smaller than $1/L$. Otherwise, the oscillatory phases cancel, and the coherent

contribution is strongly suppressed due to the sinc function

$$\mathrm{sinc}(x) = \frac{\sin x}{x}. \tag{59}$$

If the cavity coupling carries a phase

$$g_{i,m} = \tilde{g}_{i,m} e^{i\boldsymbol{k}_c\cdot\boldsymbol{R}_i}, \tag{60}$$

the integral is accordingly modified to

$$\frac{\rho^2}{N}\iint \mathrm{d}^3 r_1 \mathrm{d}^3 r_2 e^{i(\boldsymbol{k}_I-\boldsymbol{k}_S-\boldsymbol{k}_c)\cdot(\boldsymbol{r}_1-\boldsymbol{r}_2)} = \frac{\rho^2}{N}V^2 \prod_{\alpha=x,y,z} \mathrm{sinc}^2\left((\boldsymbol{k}_I-\boldsymbol{k}_S-\boldsymbol{k}_c)_\alpha \frac{L_\alpha}{2}\right) = \tag{61}$$

$$\frac{N_I^2}{N} \prod_{\alpha=x,y,z} \mathrm{sinc}^2\left((\boldsymbol{k}_I-\boldsymbol{k}_S-\boldsymbol{k}_c)_\alpha \frac{L_\alpha}{2}\right), \tag{62}$$

which introduces the cavity-coherence scale into the expression, without fundamentally altering the result. In different geometries (e.g., with curved mirrors or non-uniformly filled cavities), the structure factor will carry the details of the integration volume geometry. In any case, coherence is critical for observing a Raman signal from the polaritonic states, as the incoherent component is suppressed by the $1/N$ factor coming from the molecular component of the polariton. In addition, the number of illuminated molecules $N_I$ must be comparable to the number of molecules collectively coupled to the cavity modes $N$ forming the polaritons. However, as we show below, orientational averaging can suppress the coherent contribution even under phase-matching conditions.

### 2. *The problem of orientational averaging*

To compute the Raman intensity from the polaritonic states, we need the orientational average[1–3]

$$\langle \tilde{g}_{j,m'}(\tilde{g}_{i,m})^* \langle 1_{i,m}| \alpha^i_{pq} |0_{i,m}\rangle \langle 1_{j,m'}| \alpha^j_{kl} |0_{j,m'}\rangle^* \rangle_O . \tag{63}$$

As discussed in the previous section, coherence is fundamental for the observability of polaritonic signals, so we can restrict our discussion to $i \neq j$. Then the average for molecules $i$ and $j$ can be performed independently (considering a perfectly disordered molecular ensemble), and thus the average factorizes

$$\langle \tilde{g}_{j,m'}(\tilde{g}_{i,m})^* \langle 1_{i,m}| \alpha^i_{pq} |0_{i,m}\rangle \langle 1_{j,m'}| \alpha^j_{kl} |0_{j,m'}\rangle^* \rangle_O = \langle (\tilde{g}_{i,m})^* \langle 1_{i,m}| \alpha^i_{pq} |0_{i,m}\rangle \rangle_O \langle \tilde{g}_{j,m'} \langle 1_{j,m'}| \alpha^j_{kl} |0_{j,m'}\rangle^* \rangle_O . \tag{64}$$

Introducing again the lab frame (Latin indices) and molecular frame (Greek indices), the polarizability $\alpha_{\alpha\beta}$ and the coupling $g_{i,m} \propto (\boldsymbol{\epsilon}^{(c)} \cdot \boldsymbol{\mu}_{i,m})$ carry two and one direction cosine, respectively

$$\alpha_{ij} = l_{i\alpha}\, l_{j\beta}\, \alpha_{\alpha\beta} \tag{65}$$

$$g \propto l_{k\gamma}\, \mu_{\gamma}. \tag{66}$$

Therefore, we need the following (isotropic) average[2]

$$\langle l_{i\alpha} l_{j\beta} l_{k\gamma} \rangle = \frac{1}{6} \epsilon_{ijk} \epsilon_{\alpha\beta\gamma}. \tag{67}$$

Raman polarizability is symmetric in the Placzek approximation, so it is reasonable to expect a strongly suppressed signal as $\epsilon_{\alpha\beta\gamma}\alpha_{\alpha\beta}\mu_{\gamma}^{m} = 0$ for $\alpha_{\alpha\beta} = \alpha_{\beta\alpha}$. Beyond Placzek, the asymmetric part of the Raman scattering contributes to the Raman scattering intensity, though the signal is then expected to be much smaller than in the case of standard spontaneous Raman scattering. The problem is, however, not insurmountable, as, for instance, resonance Raman can generate significant antisymmetric contributions to the effective Raman tensor. Notice that Equation 67 further imposes geometrical constraints on the polarization of the incident, scattered, and cavity light from the Levi-Civita tensor $\epsilon_{ijk}$.

The cavity design can, nevertheless, lift the problem of intensity suppression due to orientational averaging. If the light-matter coupling were approximately uniform $g_i \to g_0$,[4] the couplings would not be present in the average expression, and the signal would then become proportional to the trace of the Raman polarizability $\alpha_{\alpha\alpha}$. The approximation of a quasi-uniform cavity coupling is not as severe a requirement as it could appear at first sight, since recent effective theories suggest that the cavity dresses quasi-uniformly the molecules.[6] The model for the uniform coupling $g_0$ will depend on the details of the cavity geometry and the model assumptions. For the present, we assume $g_0$ to be proportional to $\partial|\boldsymbol{\mu}_{i,m}|/\partial Q_m$, which is rotationally invariant, with the same dimension as the transition dipole between two vibrational states, and fixes a consistent sign convention for the normal modes and the molecular states.[4]

Finally, non-isotropic systems will, in general, not lead to zero orientational averaging of the single-molecule contribution.

In summary, for an isotropic molecular ensemble and a symmetric Placzek Raman tensor, orientational averaging suppresses the coherent polaritonic Raman amplitude even when

the cavity coupling still depends on molecular orientation. Observable polaritonic Raman scattering therefore requires at least one of the following

- Approximately uniform cavity dressing $g_{i,m} \to g_m^0$

- Partial orientational ordering or sample anisotropy

- Antisymmetric Raman tensor components beyond the Placzek approximation (e.g. Resonance Raman).

## III. COMPUTATIONAL DETAILS

The simulations of the normal modes, transition dipoles, and Raman polarizabilities are obtained from the $e^T$ program,[7] an open-source electronic-structure program. All the calculations are performed at Hartree-Fock level of theory, using a 6-31g* basis set. The initial molecular structures are reported in section III A. The geometries are then optimized to the ground state minimum $\boldsymbol{Q}_0$, and the nuclear Hessian is computed from numerical differentiation of the nuclear gradients. Molecular modes are visualized using OVITO.[8] The Raman scattering intensities are computed in the Placzek approximation, assuming a uniform cavity coupling as explained in section II B 2. The intensity is thus proportional to the fourth power of the scattered frequency and the polaritonic Raman scattering activity $S_p$ (atomic units are assumed, and Boltzmann factors are omitted since we exclusively model Stokes radiation assuming low temperature $T$)

$$I_R \propto S_p(\omega_I - \omega_p)^4, \tag{68}$$

where $\omega_p$ is the energy of the polariton state, and $\omega_I$ is the energy of the incident laser, here assumed as a Nd:YAG laser source operating at 1064 nm wavelength. According to the derivation in section II for a uniform cavity dressing, the Raman scattering activity here is (using the notation of section I and section II)

$$S_p = \left| \sum_m \frac{v_m^p}{\sqrt{2\omega_m}} \frac{\partial \alpha_{\alpha\alpha}^e}{\partial \boldsymbol{Q}_m} \bigg|_{\boldsymbol{Q}=\boldsymbol{Q}_0} \right|^2 . \tag{69}$$

The phase of the different molecular vibrations is fixed by assuming $g_m^0 \propto \partial|\boldsymbol{\mu}|/\boldsymbol{Q}_m$ as explained in section II B 2.[4] All the spectra are broadened with a Lorentzian broadening according to the width function

$$F(\omega) = \frac{1}{\pi}\frac{\gamma}{(\omega-\omega_p)^2+\gamma^2}, \tag{70}$$

and the applied broadenings is $\gamma = 15$ cm$^{-1}$, typical of molecular vibrations. All the spectra are normalized such that the maximum intensity is 1 (arbitrary units). The (collective) light matter coupling strength is set to achieve Rabi splittings of the order of $30-100$ cm$^{-1}$, as typical of experiments of vibrational strong coupling. For tryptophan, the mode $m_1$ (see Figure 1 in the main text) has a collective coupling strength of $\sim 19$ cm$^{-1}$ (all the other modes scale accordingly with their transition dipole moment). For histidine, the mode at 934 cm$^{-1}$ has a light-matter coupling of $\sim 31$ cm$^{-1}$. Finally, for proline, the off-resonant modes at 3252 and 3261 cm$^{-1}$ have a coupling of $\sim 19$ cm$^{-1}$, corresponding to a perturbative parameter $g/\Delta \sim 0.20$ and an effective coupling $g_{eff} = g^2/\Delta \sim 5$ cm$^{-1}$ which, although small, is comparable to the mode splitting $\sim 10$ cm$^{-1}$ and thus induces relevant mixing. While $g/\Delta \sim 0.20$ might be too large for a reliable examination based on the second-order perturbation theory, the analysis is qualitatively valuable. All input, output, and script files are available from the following Zenodo repository: 10.5281/zenodo.20625724.

### A. Molecular geometries

Table I. Initial molecular geometry of tryptophan. The coordinates are provided in Angstrom Å.

| Atom | x [Å] | y [Å] | z [Å] |
|---|---|---|---|
| O | 1.4762 | -0.8154 | 1.7282 |
| O | 2.5598 | -1.9472 | 0.0879 |
| N | -1.6474 | 2.0044 | 0.5482 |
| N | 3.8759 | 0.3655 | -0.5832 |
| C | 0.1475 | 1.1323 | -0.4236 |
| C | 1.5065 | 0.9845 | -0.9830 |
| C | -0.8667 | 0.1349 | -0.3859 |
| C | 2.5601 | 0.4881 | 0.0402 |
| C | -1.9823 | 0.7078 | 0.2286 |
| C | -0.3625 | 2.2702 | 0.1579 |
| C | -0.9431 | -1.2045 | -0.8169 |
| C | -3.1751 | 0.0055 | 0.4346 |
| C | 2.2083 | -0.8852 | 0.5860 |
| C | -2.1298 | -1.9193 | -0.6176 |
| C | -3.2274 | -1.3216 | -0.0012 |
| H | 1.4724 | 0.2941 | -1.8370 |
| H | 1.8346 | 1.9497 | -1.3927 |
| H | 2.6391 | 1.1902 | 0.8789 |
| H | 0.0723 | 3.2449 | 0.3304 |
| H | -2.2620 | 2.6639 | 1.0051 |
| H | -0.1001 | -1.6864 | -1.3035 |
| H | -4.0301 | 0.4699 | 0.9143 |
| H | -2.1948 | -2.9522 | -0.9490 |
| H | 4.5715 | 0.1348 | 0.1255 |
| H | 4.1549 | 1.2657 | -0.9717 |
| H | -4.1393 | -1.8953 | 0.1434 |
| H | 1.2617 | -1.7057 | 2.0797 |

Table II. Initial molecular geometry of histidine. The coordinates are provided in Angstrom Å.

| Atom | x [Å] | y [Å] | z [Å] |
|---|---|---|---|
| O | -1.0689 | -1.6919 | -0.7660 |
| O | -2.4946 | -1.1223 | 0.9038 |
| N | 1.6797 | 0.1409 | 1.1624 |
| N | -2.7494 | 1.3693 | -0.2178 |
| N | 2.8201 | -0.3365 | -0.6825 |
| C | -0.3579 | 1.3004 | 0.3217 |
| C | -1.5062 | 0.6397 | -0.4567 |
| C | 0.9671 | 0.6020 | 0.1195 |
| C | -1.7498 | -0.7917 | -0.0102 |
| C | 1.6556 | 0.3150 | -1.0031 |
| C | 2.8043 | -0.4249 | 0.6290 |
| H | -0.5942 | 1.3196 | 1.3946 |
| H | -0.2433 | 2.3450 | 0.0039 |
| H | -1.3124 | 0.6416 | -1.5355 |
| H | 1.4441 | 0.1943 | 2.1450 |
| H | 1.3992 | 0.5299 | -2.0301 |
| H | -2.6419 | 2.3346 | -0.5280 |
| H | -2.9358 | 1.4198 | 0.7835 |
| H | 3.5655 | -0.8810 | 1.2458 |
| H | -1.2415 | -2.6142 | -0.4801 |

Table III. Initial molecular geometry of proline. The coordinates are provided in Angstrom Å.

| Atom | x [Å] | y [Å] | z [Å] |
|---|---|---|---|
| N | 0.859959 | 1.091243 | -0.169345 |
| C | -0.000904 | -0.059810 | -0.508705 |
| H | 0.001976 | -0.215965 | -1.594691 |
| C | -1.444715 | 0.179531 | -0.076962 |
| O | -2.012458 | 1.255461 | 0.037798 |
| O | -2.149899 | -0.963009 | 0.069292 |
| C | 0.686038 | -1.223190 | 0.199248 |
| C | 2.125106 | -0.765946 | 0.423636 |
| C | 2.221574 | 0.584942 | -0.261549 |
| H | 0.689782 | 1.866436 | -0.808342 |
| H | -3.046256 | -0.644235 | 0.306442 |
| H | 0.219842 | -1.470386 | 1.160115 |
| H | 0.658506 | -2.124687 | -0.422986 |
| H | 2.315388 | -0.665891 | 1.499124 |
| H | 2.849310 | -1.480572 | 0.020130 |
| H | 2.520106 | 0.469486 | -1.310075 |
| H | 2.938147 | 1.246292 | 0.233471 |

## REFERENCES


[1] L. D. Barron, *Molecular light scattering and optical activity* (Cambridge University Press, 2009).

[2] D. P. Craig and T. Thirunamachandran, *Molecular quantum electrodynamics* (Courier Corporation, 2012).

[3] S. S. Andrews, Using rotational averaging to calculate the bulk response of isotropic and anisotropic samples from molecular parameters, Journal of chemical education **81**, 877 (2004).

[4] In writing $g_{i,m} \equiv g_m^0$, one implicitly assumes that all couplings share the same phase. In general, however, the coupling constants

$$g_{i,m} \propto e^{i\mathbf{k}_c \cdot \mathbf{R}_i} \left( \boldsymbol{\epsilon}^{(c)} \cdot \boldsymbol{\mu}_{i,m} \right)$$

are complex and may differ by phase factors arising both from molecular positions and from the (arbitrary) phase convention of the vibrational transition dipole moments. This phase freedom is not physical. A local phase transformation of the form

$$a_{i,m} \to e^{i\phi_i} a_{i,m}$$

can always be used to absorb the phase of $g_{i,m}$, rendering the couplings real and positive. Physical observables depend only on gauge-invariant quantities such as $|g_{i,m}|^2$, and are therefore unaffected by this choice. The assumption $g_{i,m} \equiv g_m^0$ thus corresponds to fixing a convenient sign in which all couplings are phase-aligned, leading to the symmetric bright mode in Eq. (27). Such a choice is arbitrary but does not constitute a fundamental physical problem, as the same phase will automatically be fixed for all other (transition) properties of the system.

[5] B. N. Parlett, *The symmetric eigenvalue problem* (SIAM, 1998).

[6] M. Kamper Svendsen, M. Ruggenthaler, H. Hübener, C. Schäfer, M. Eckstein, A. Rubio, and S. Latini, Theory of quantum light-matter interaction in cavities: Extended systems and the long wavelength approximation, arXiv e-prints , arXiv (2023).

[7] S. D. Folkestad, E. F. Kjønstad, A. C. Paul, R. H. Myhre, R. Alessandro, S. Angelico, A. Balbi, A. Barlini, A. Bianchi, C. Cappelli, *et al.*, et 2.0: An efficient open-source molecular electronic structure program, The Journal of Chemical Physics **164** (2026).

[8] A. Stukowski, Visualization and analysis of atomistic simulation data with ovito–the open visualization tool, Modelling and simulation in materials science and engineering **18**, 015012 (2010).


\end{widetext}

\end{document}